# Inverse-Designed Nanobeam Cavities for High-Cooperativity Spin–Photon Interfaces in Silicon

Naïl Letty[1], Md. Sakibul Islam[1], Wayesh Qarony[1,2,3,*]

[1]Department of Electrical and Computer Engineering, University of Central Florida, Orlando, FL 32816, USA

[2]Department of Physics, University of Central Florida, Orlando, FL 32816, USA

[3]CREOL, The College of Optics and Photonics, University of Central Florida, Orlando, FL 32816, USA

*Corresponding Email: wayesh@ucf.edu

## Abstract

Telecom-band spin defects in silicon offer a compelling platform for integrated quantum networks, combining silicon photonic processing with spin-based quantum memories and multiqubit registers for heralded entanglement between network nodes. The *T* center and recently demonstrated *Al1* center are particularly promising, but their limited coherent zero-phonon emission makes strong cavity enhancement essential for efficient spin–photon interfaces. Here, we inverse-design silicon nanobeam cavities for both centers using a unified adjoint-optimization framework that directly targets Purcell enhancement, resonance alignment, and photon extraction. The framework produces two complementary cavity architectures for each emitter: a symmetric cavity optimized for maximal Purcell enhancement and an asymmetric cavity engineered for directional coupling to an integrated waveguide. Three-dimensional finite-difference time-domain simulations yield loaded quality factors up to $3.25\times10^6$, Purcell factors up to $7.49\times10^5$, and cooperativities of approximately $1.4\times10^3$ to $2.6\times10^4$, corresponding to cavity-coupled emission fractions exceeding 99.9%. At the nominal design, the asymmetric cavities direct approximately 90% of the emitted power into a single on-chip waveguide while maintaining cooperativity above $1.4\times10^3$. Under nanometer-scale fabrication perturbations, all four designs remain within the high-cooperativity regime, and high cooperativity persists when the cavity quality factor is constrained to a realistic absorption-limited value of $10^5$. By jointly optimizing resonance, Purcell enhancement, and directional extraction within a single fabrication-compatible parametrization, this work provides a unified cavity-design route for both of silicon's telecom spin centers, and, to our knowledge, the first cavity designs for the *Al1* center.

**Keywords:** *Silicon photonics, spin–photon interface, color centers, inverse design, nanobeam cavity, cavity QED, Purcell enhancement*

## 1. Introduction

A quantum network requires efficient interfaces between stationary quantum memories and flying photonic qubits, enabling quantum information stored in a spin to be transferred to photons and distributed between remote nodes.[1-4] The performance of such a spin–photon interface is governed not simply by photon brightness, but by how efficiently the emitter couples to a coherent, collectable optical mode. This interaction is quantified by the cooperativity $C$, which compares coherent emitter–photon coupling with competing loss channels. In the high-cooperativity regime ($C$>>1), emission is preferentially directed into the desired optical mode and the error of spin–photon operations decreases approximately as $1/C$. Achieving high cooperativity is therefore a central requirement for scalable spin–photon quantum interfaces.

Silicon provides an attractive platform for realizing such interfaces by combining telecom-band optical transitions, long-lived electronic and nuclear spin degrees of freedom, and compatibility with mature CMOS and silicon-photonics processing.[5-7] This combination creates a pathway toward integrating quantum memories, multiqubit spin registers, photonic routing, and single-photon detection within a common material platform. Recent advances have further enabled deterministic or spatially controlled formation of silicon quantum emitters[8,9] and high-performance telecom single-photon detection compatible with integrated photonic circuits.[10,11] While spin–photon interfaces are also advancing rapidly in other material systems, including industrially manufactured III–V quantum dots[12,13], color centers in diamond and SiC[14-17], as well as two-dimensional materials like *hBN*[18], silicon offers the distinctive prospect of embedding optically addressable spins directly within a scalable photonic-processing platform.

Among silicon color centers, and as distinct from rare-earth dopants such as erbium, whose intra-4f transitions form a separate class of telecom spin–photon interface, the *T* center and the recently demonstrated *Al1* center are particularly promising for telecom-band spin–photon interfaces. The *T* center, a carbon–hydrogen complex emitting near 1326 nm, has progressed from single-emitter isolation[19] to coherent control of its electron and nuclear spins[20] and entangled multiqubit registers integrated with silicon waveguides.[21] These capabilities provide the essential ingredients for spin-based quantum memories and heralded entanglement between photonic-network nodes. The *Al1* center, an aluminum–carbon complex emitting at 1482.44 nm, has recently been demonstrated at the single-emitter level and likewise possesses an addressable ground-state spin.[22] Its reported optical emission is approximately twenty times brighter than that of the *T* center, making it an attractive complementary candidate for telecom spin–photon interfaces, with cavity enhancement identified as an important next step toward spin–photon entanglement.[22] Together, these centers provide compelling targets for cavity-integrated silicon quantum photonics.

A fundamental limitation, however, is that only a fraction of their optical emission occurs through the coherent zero-phonon line (ZPL). The corresponding Debye–Waller factors are approximately 0.23 for the *T* center[20] and 0.14 for the *Al1* center,[23] the latter obtained for the aluminum–carbon defect among the *T* center-like family in that first-principles study, while emission from a dipole embedded in high-index silicon ($n \approx 3.46$) is also poorly matched to a single collectable photonic mode. As bare emitters in bulk silicon, without a cavity, these centers couple only weakly to any single collectable mode, well below the $C \gg 1$ regime, which a resonant cavity must overcome. Consequently, efficient spin–photon interfacing requires simultaneous enhancement of the coherent ZPL transition and control of photon extraction. Recent analysis indicates that approximately $10^2$ - $10^3$-fold Purcell enhancement may be required to approach a Fourier-limited, network-compatible silicon spin–photon interface.[24] A resonant nanophotonic cavity provides a direct route to this regime by enhancing spontaneous emission into the ZPL through the Purcell effect while redirecting the emitted photons into a well-defined optical mode.[25]

Conventional photonic-crystal nanobeam cavities can achieve high-quality factors and small mode volumes through analytical or deterministic design approaches.[26,27] However, a quantum-network interface requires several competing properties simultaneously: resonance with a prescribed emitter wavelength, large $Q/V_{\mathrm{eff}}$, strong emitter–cavity coupling, and efficient photon extraction into a useful optical channel. Inverse design provides a natural framework for this multidimensional problem because the cavity geometry can be optimized directly against the desired electromagnetic figure of merit through adjoint gradients.[28] Such approaches have become increasingly powerful for nanophotonic optimization[29,30] and have been extended to the design of solid-state single-photon sources with simultaneous control of emission enhancement and photon extraction.[31] Here we do not pursue free-form topology optimization; instead, adjoint gradients tune only a compact set of lattice parameters of an otherwise conventional nanobeam, so that the advantage of the approach lies in jointly satisfying the resonance, enhancement, and extraction objectives rather than in the geometric freedom of the design space.

Here, building on our silicon nanophotonic platform,[32] we develop a unified adjoint inverse-design framework for nanobeam cavities targeting both the *T* and *Al1* centers. A compact parametrization based only on the photonic-crystal lattice constants and a single hole fill factor produces two complementary cavity architectures for each emitter: a symmetric cavity optimized for maximal

Purcell enhancement and an asymmetric cavity optimized for directional coupling into a single on-chip waveguide. Three-dimensional FDTD simulations show that all four designs reach the high-cooperativity regime while maintaining compact footprints, with the asymmetric cavities combining strong cavity enhancement with approximately 90% directional photon extraction at the nominal design. We further evaluate their robustness to nanometer-scale fabrication disorder and realistic material-loss limits and benchmark the resulting designs against reported silicon color-center cavities. The central result is that one adjoint pipeline, built on a single fabrication-compatible parametrization, produces both a maximal-Purcell and a directional cavity for each of silicon's two telecom spin centers, with all four designs reaching the high-cooperativity regime. To our knowledge, these are the first cavity designs to target the *Al1* center.

## 2. Results and Discussion

### *2.1. Inverse-designed cavities for the T and Al1 centers*

We develop a unified inverse-design framework for nanobeam cavities resonant with the *T*- and *Al1*-center zero-phonon lines. Each cavity consists of a one-dimensional silicon nanobeam patterned in a 220-nm-thick, 500-nm-wide silicon-on-insulator device layer and released to form an air-clad membrane. The 500 nm beam width is chosen so that the beam supports a single guided mode at the target wavelengths in the 220 nm device layer and matches the standard silicon-photonic waveguide cross-section, keeping the geometry compatible with established silicon fabrication. A sequence of air holes defines the photonic crystal, while a gradual variation of the lattice period toward the cavity center forms the central taper that localizes the optical mode at the position of the color center. Rather than allowing arbitrary boundary deformation, we deliberately restrict the design to a small set of physically meaningful parameters: the taper and mirror lattice constants and a single fill factor *FF* = r/a, which determines each hole radius from its local period. This low-dimensional parametrization retains the established nanobeam-cavity architecture [33,34] while enabling direct optimization of its cavity-QED performance and preserving a geometry that can be readily translated to a fabrication mask.

The inverse-design pipeline is illustrated in Fig. 1a. The design vector, P = (*FF*, $a_{mir}$, $a_0$, ..., $a_4$) contains the fill factor, mirror period, and five taper periods. For each trial geometry, a forward FDTD simulation determines the cavity resonance and the relevant optical figure of merit, while an adjoint simulation evaluates its gradient with respect to P. A sign-based gradient-ascent update (Eq. S2) then advances the geometry, requiring only one forward and one adjoint simulation per iteration. Two cavity functionalities are obtained within the same framework. The symmetric cavity uses seven mirror periods on each side of the taper and is optimized for maximal Purcell enhancement. In the asymmetric cavity, the right mirror is reduced to three periods while seven periods are retained on the left, intentionally introducing an output channel for preferential photon extraction into the right-hand waveguide.

To simultaneously optimize cavity enhancement, photon extraction, and spectral alignment, we maximize the objective, in which the Purcell factor $F_P$, defined in Equation (2), quantifies the cavity-enhanced spontaneous-emission rate of a dipole at the field maximum, and $\beta_R$ is the fraction of emitted power directed into the right-hand output

$$O = R\exp\left(-\frac{\delta^2}{2\sigma^2}\right) - \mu\,\delta^2 \quad (1)$$

Where $\delta = (\lambda_{res} - \lambda_0)/\lambda_0$ is the fractional detuning from the target color-center transition, $\sigma$ sets the width of the Gaussian spectral acceptance, and μ weights the quadratic detuning penalty (values in Section S1). For the symmetric cavity, the reward is $R = F_P$, whereas for the asymmetric cavity it becomes $R = F_P \cdot \beta_R$. The latter objective therefore balances strong cavity enhancement with directional photon extraction rather than maximizing the Purcell factor alone, following the broader concept of multi-objective inverse design for integrated single-photon sources.[31] The Gaussian spectral weighting and quadratic detuning penalty constrain the resonance to the target transition while the optical reward is maximized.

The optimization trajectories demonstrate this simultaneous spectral and cavity optimization. For *AI1*, the resonance evolves toward 1482.44 nm as the quality factor increases and subsequently stabilizes (Fig. 1c); the corresponding *T*-center cavity converges to 1326 nm with similar behavior (Fig. 1d). In both cases, convergence occurs within approximately 10–20 iterations. Each cavity is produced by a single adjoint optimization initialized from an analytically designed nanobeam of the target architecture, and, as is standard for non-convex nanophotonic inverse design, the procedure converges to a locally optimized geometry rather than a certified global optimum.[35-37] The quality factors plotted during optimization are evaluated at the optimization resolution and therefore differ from the converged high-resolution values reported below (Computational Section). At convergence, the simulated mode intensity $|E|^2$ is strongly localized at the taper center for both emitters (Fig. 1e,f), placing the field maximum at the intended color-center position.[8] Figure 1 presents the asymmetric cavities as representative optimization trajectories; the corresponding symmetric-cavity convergence is provided in Fig. S1.

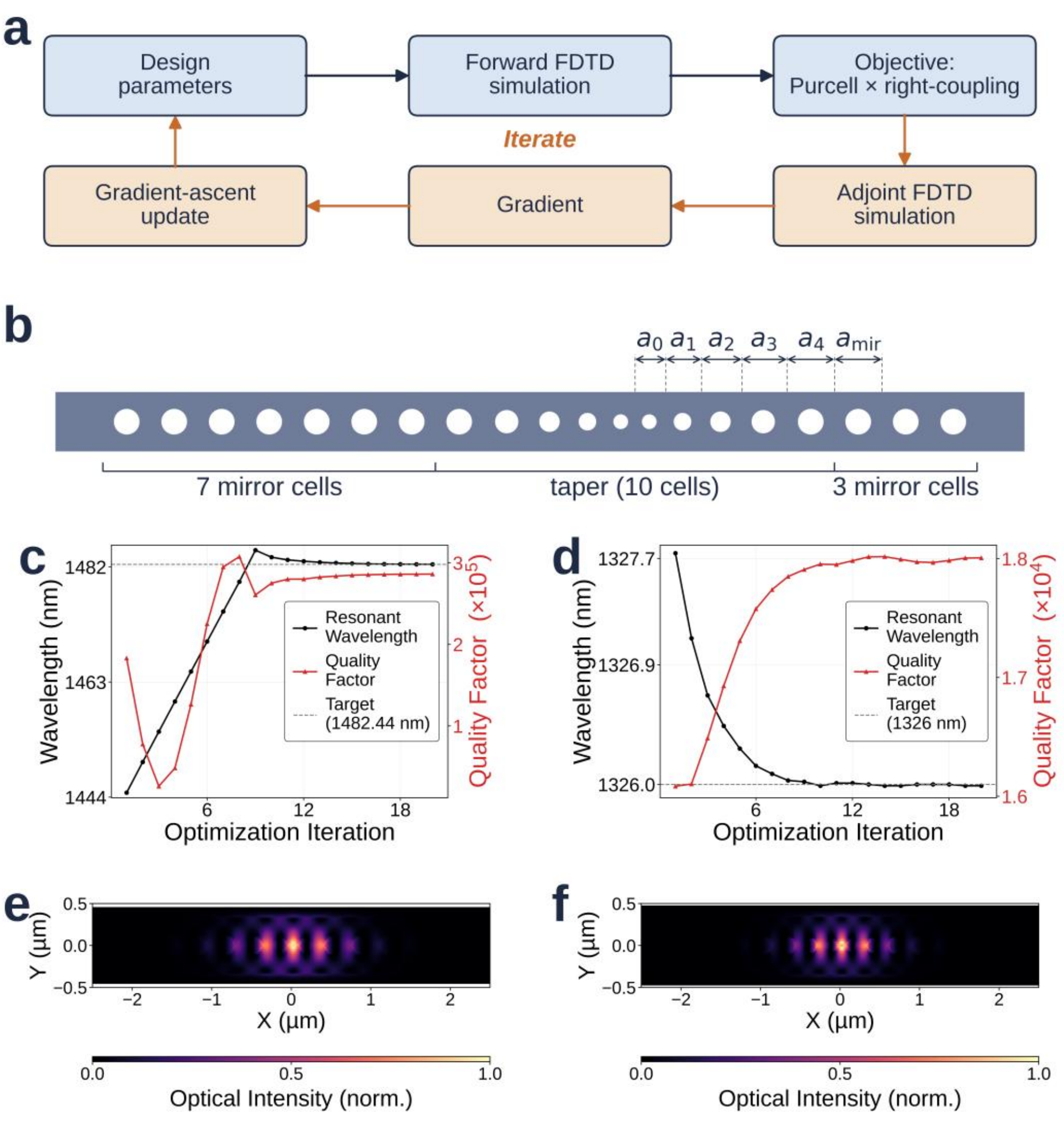

***Figure 1. Inverse-design framework, cavity geometry, and optimization convergence for the T- and Al1-center nanobeam cavities. (a) Adjoint-optimization loop in which a forward FDTD simulation evaluates the cavity response and objective, an adjoint simulation computes the gradient with respect to the design parameters, and gradient ascent iteratively updates the cavity geometry. The objective maximizes the Purcell factor $F_P$ for the symmetric cavity and $F_P \cdot \beta_R$ for the asymmetric cavity, where $\beta_R$ is the right-hand waveguide-coupling fraction. (b) Schematic of the nanobeam photonic crystal, showing the mirror period $a_{mir}$ and the five taper periods $a_0$ to $a_4$, together with the seven mirror cells, the central taper, and the shortened three-cell right mirror of the asymmetric design. (c,d) Optimization trajectories for the (c) Al1 and (d) T cavities, showing convergence of the resonance toward the target wavelengths of 1482.44 and 1326 nm, respectively, while the quality factor increases and stabilizes. The plotted quality factors are evaluated at the***

***optimization resolution and therefore differ from the converged high-resolution values reported in Table 1. (e,f) Normalized mode intensity |E|$^2$ of the converged (e) Al1 and (f) T asymmetric cavities, showing localization at the taper center; the intensity appears nearly symmetric because the right-directed leakage that provides the directional out-coupling is small compared with the central antinode. The asymmetric cavity family is shown as the representative case; the corresponding symmetric-cavity convergence is provided in Figure S1.***

### *2.2. Balancing cavity enhancement and directional photon extraction*

The same inverse-design framework produces two complementary cavity architectures that address different requirements of an integrated spin–photon interface (Fig. 2). The symmetric cavity uses balanced mirrors to maximize optical confinement and Purcell enhancement. With no preferential in-plane output channel, approximately 87% of the emitted power in the *Al1* design leaves transversely to the nanobeam axis through the top, bottom, front, and back boundaries, while only a small fraction exits through the end waveguides. The asymmetric cavity instead deliberately weakens one mirror, over-coupling the cavity mode to a single output waveguide and thereby converting the confined cavity field into a directional on-chip photon channel, analogous to nanobeam cavities intentionally over-coupled to an access waveguide.[34] For the *Al1* design, approximately 90% of the emitted power is directed into the right-hand waveguide.

The distinct extraction behavior is evident in the three-dimensional mode-field renderings of Figs. 2c,d (raw simulated field maps in the insets). The symmetric cavity exhibits a strongly confined standing-wave mode centered at the taper, whereas the asymmetric cavity retains localization near the emitter while extending the field preferentially toward the weakened right mirror. A per-boundary decomposition of the radiated power quantifies this redistribution (Fig. 2e,f): the symmetric cavity spreads its emission almost equally over the four transverse faces, with only about 7% leaving through the right-hand waveguide, whereas the asymmetric cavity concentrates the flux into the right output face, reaching a directional fraction of 0.90 for *Al1* and 0.93 for *T*; the field antinodes lie in the silicon between the air holes, confirming the dielectric character of the mode. The full six-face decomposition of all four cavities is provided in Section S5 and Figure S4. Thus, mirror asymmetry provides a controlled mechanism for trading some cavity confinement for efficient delivery of cavity-enhanced photons into a single on-chip waveguide.

Directional photon extraction is a key requirement for integrating color-center cavities into larger photonic circuits. Related approaches have pursued near-unity emitter-to-fiber interfacing through tailored photonic-crystal profiles [38] and controlled far-field emission through inverse design.[39] Here, both operating regimes emerge from the same compact cavity parametrization: the symmetric architecture prioritizes maximal Purcell enhancement, whereas the asymmetric architecture jointly optimizes cavity enhancement and directional waveguide coupling. The design choice can therefore be set by the intended function of the spin–photon interface – maximal light–matter interaction or efficient routing of cavity-enhanced photons into an integrated quantum photonic circuit.

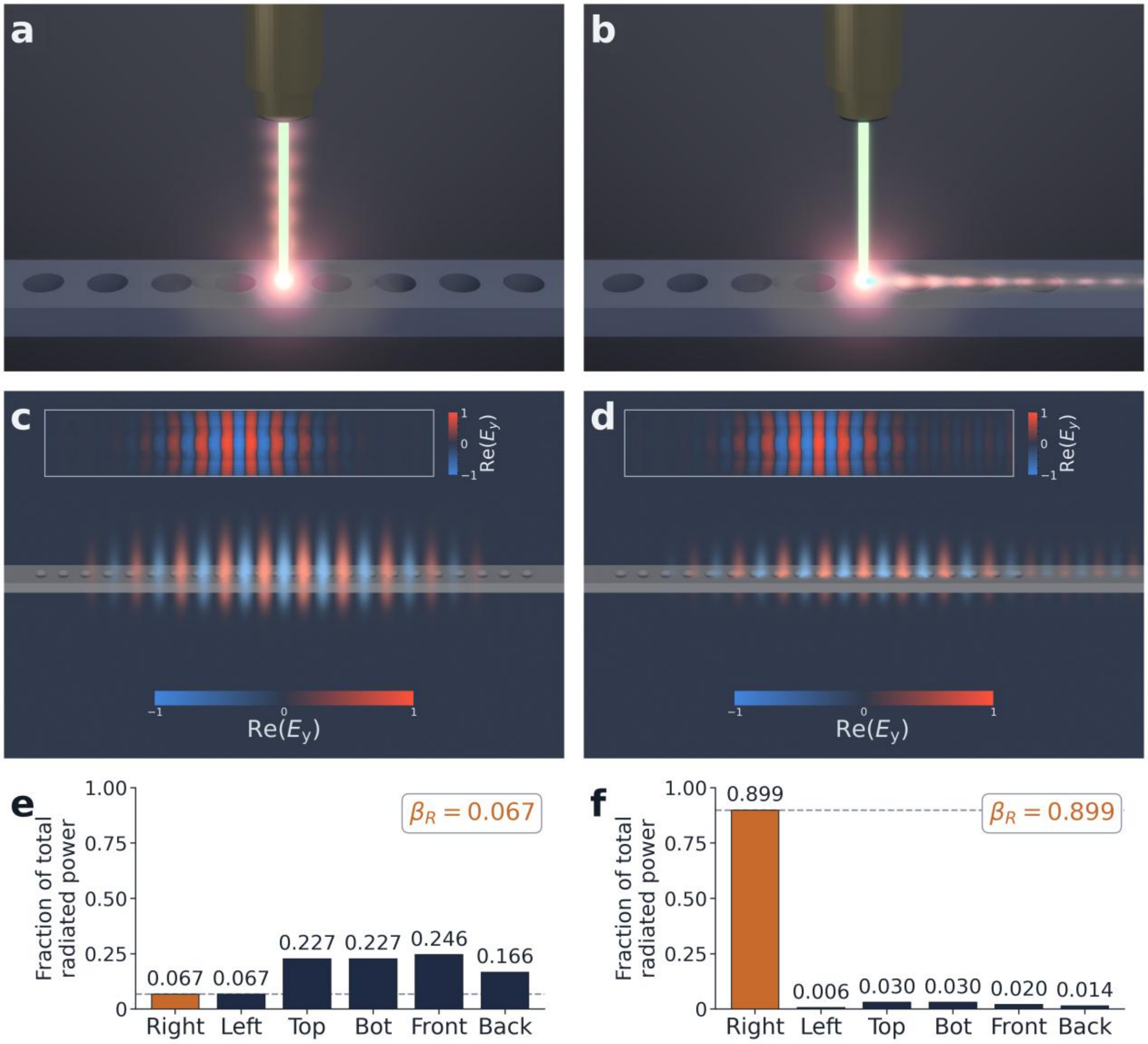

**Figure 2. Complementary cavity architectures for enhanced light–matter interaction and directional photon extraction. (a,b) Operating principle of the (a) symmetric and (b) asymmetric nanobeam cavities. The vertical beam denotes out-of-plane optical excitation**

**of the embedded emitter: the symmetric cavity re-radiates predominantly out of plane, whereas the asymmetric cavity redirects the emission into the right-hand waveguide, consistent with the per-face flux in (e,f). The symmetric cavity uses balanced mirrors to maximize optical confinement, with approximately 87% of the emitted power in the *AI1* design leaving transversely to the nanobeam axis. The asymmetric cavity uses deliberately unbalanced mirrors to preferentially couple the cavity emission into a single on-chip waveguide, reaching approximately 90% right-directed emission for *AI1*. (c,d) Mode-field visualization of the (c) symmetric and (d) asymmetric *AI1* cavities: a three-dimensional rendering of the nanobeam in which the simulated in-plane field Re($E_y$) is mapped onto brightness, shown with the raw FDTD field map of the same mode in the inset (shared color scale). The rendering conveys the confined standing-wave mode of the symmetric cavity and the preferential right-hand out-coupling of the asymmetric cavity; the field antinodes sit in the silicon between the air holes, consistent with a dielectric-band mode. (e,f) Per-face decomposition of the radiated power for the (e) symmetric and (f) asymmetric *AI1* cavities: the fraction of the total emitted power leaving each of the six nanobeam faces, with the right output face highlighted and its value $\beta_R$ quoted. The full four-cavity decomposition is provided in Figure S4.**

### *2.3. Cavity-QED figures of merit*

The performance of the inverse-designed cavities as spin–photon interfaces is quantified by the Purcell factor, cooperativity, and cavity-coupled emission fraction. For a resonant dipole positioned at the cavity-field maximum, the Purcell factor is

$$F_P = \frac{3}{4\pi^2}\left(\frac{\lambda}{n}\right)^3 \frac{Q}{V_{\mathrm{eff}}} \tag{2}$$

Where $Q$ is the loaded cavity quality factor and $V_{eff}$ is the effective mode volume. While $F_P$ quantifies the photonic enhancement provided by the cavity, the effective interaction with a specific color center must also account for the fraction of radiative emission in the zero-phonon line and for nonradiative decay. We therefore define the cooperativity and cavity-coupled emission fraction as

$$C = F_P \cdot \mathrm{DW} \cdot \eta_{\mathrm{QE}} \tag{3}$$

$$\beta_{\mathrm{cav}} = \frac{C}{C+1} \tag{4}$$

where $DW$ is the Debye–Waller factor and $\eta_{QE}$ is the emitter quantum efficiency. In this description, $C$>>1 corresponds to the regime in which cavity-enhanced ZPL emission dominates competing decay channels, making high cooperativity the relevant figure of merit for an efficient spin–photon interface.[25] This is the standard cavity-QED cooperativity $C = 4g^2/(\kappa\gamma)$ specialized to a color center, for which only the fraction DW·$\eta_{QE}$ of the total decay is coherent zero-phonon radiation (Section S2). $\beta_{cav}$ quantifies the fraction of emission funneled into the cavity mode and is distinct from the directional waveguide fraction $\beta_R$ (Eq. (S8)): a cavity can reach $\beta_{cav} \rightarrow 1$ while routing only part of that emission into the output waveguide.

The four optimized cavities span loaded quality factors from $1.16\times10^5$ for the asymmetric *T*-center cavity to $3.25\times10^6$ for the symmetric *Al1* cavity, while maintaining similar mode volumes of 0.324-0.332 $(\lambda/n)^3$ (Table 1). These mode volumes represent a deliberate high-*Q* operating point rather than minimization of $V_{eff}$ alone.[40] The resulting Purcell factors range from $2.7\times10^4$ to $7.5\times10^5$ (Table 1). After incorporating the emitter-specific $DW$ and $\eta_{QE}$, the predicted cooperativities remain exceptionally high: $C \approx 2.1\times10^4$ to $2.6\times10^4$ for the symmetric cavities and ~$1.4\times10^3$ to $1.8\times10^3$ for the asymmetric cavities. Consequently, all four designs reach $\beta_{cav}$>0.999 and lie deep in the saturated high-cooperativity regime (Fig. 3a).

The reduction in cooperativity from the symmetric to asymmetric cavities originates predominantly from the deliberately reduced loaded *Q*, since their mode volumes remain nearly unchanged. This is the central trade-off between the two architectures: weakening one mirror sacrifices approximately an order of magnitude in cooperativity but provides a directional photonic output. Importantly, this trade-off has little impact on $\beta_{cav}$, because both architectures remain deep in the high-cooperativity regime ($C$ >>1). The asymmetric cavities therefore combine $\beta_{cav}$ >0.999 with right-directed waveguide fractions of $\beta_R$=0.90 for *Al1* and 0.93 for *T*, simultaneously providing strong cavity-mediated emission and efficient on-chip photon extraction.

This reduction is channel-specific. Decomposing the loaded quality factor by radiation channel through the per-face power fractions, and grouping the two end faces into a waveguide-coupling quality factor $Q_{wg}$ and the four transverse faces into a scattering quality factor $Q_{scat}$ (Table S2b), the scattering channel remains essentially unchanged between the symmetric and asymmetric designs of each center ($Q_{scat}$ within a factor of 1.2 to 1.6), whereas the waveguide-coupling channel collapses by roughly two orders of magnitude ($Q_{wg}$ falls by a factor of 53 for *T* and 77 for *Al1*). The reduction in loaded *Q* from the symmetric to the asymmetric cavities is therefore governed by the intended waveguide coupling rather than by any increase in intrinsic scattering loss, so the weakened mirror acts as an output port and not as a source of excess loss.

The effective mode volumes are nearly equal between the symmetric and asymmetric cavities of each center (Table 1), because $V_{eff} = \int\varepsilon|E|^2dV / \max(\varepsilon|E|^2)$ is set by the tightly confined field at the central taper, which is common to both families. Although the weakened right mirror does allow the mode to extend further toward the output, with the in-plane field amplitude at the right monitor edge exceeding that at the left by more than an order of magnitude in the asymmetric cavities, this leaked field remains below $10^{-4}$ of the peak energy density and contributes less than 0.1% of the integral. This reflects the standard behavior of photonic-crystal nanobeam cavities, in which the effective mode volume stays wavelength-scale and is fixed by the central taper, while the quality factor is controlled by the mirror strength, so the two are largely decoupled.[27] A convergence check of $V_{eff}$ against the integration-box size shows a clean plateau for boxes longer than 3 μm (Figure S3), and the symmetric and asymmetric values agree to within 1.2% for both centers. The resulting change in the coherent coupling rate, $g \propto 1/\sqrt{V_{eff}}$, is below 0.6%, so all cooperativities are unchanged at the level relevant to the $C \gg 1$ conclusion.

***Table 1.*** *Simulated cavity-QED figures of merit for the four inverse-design nanobeam cavities obtained from high-resolution FDTD calculations. Cooperativity C and cavity-coupled emission fraction $\beta_{cav}$ are calculated using the emitter parameters described in Section S3 (T: DW = 0.23, $\eta_{QE}$ = 0.234; AI1: DW = 0.14, $\eta_{QE}$ = 0.20); the coupling-to-loss ratio 2g/κ uses $Q_{crit} \approx 1.6\times10^5$ for AI1 and ≈ $3.4\times10^5$ for T. Additional metrics are provided in Table S2a.*

| Device | $\lambda_{res}$ (nm) | Q (loaded) | $V_{eff}$ $((\lambda/n)^3)$ | $F_P$ | C | $\beta_{cav}$ | 2g/κ | $\beta_R$ | Footprint ($\mu m^2$) |
|---|---|---|---|---|---|---|---|---|---|
| *AI1* symmetric | 1482.26 | $3.25\times10^6$ | 0.330 | $7.49\times10^5$ | $2.10\times10^4$ | 0.99995 | 20 | 0.067 | 4.34 |
| *AI1* asymmetric | 1482.32 | $2.87\times10^5$ | 0.332 | $6.58\times10^4$ | $1.84\times10^3$ | 0.99946 | 1.8 | 0.899 | 3.54 |
| *T* symmetric | 1325.45 | $2.02\times10^6$ | 0.324 | $4.74\times10^5$ | $2.55\times10^4$ | 0.99996 | 6.0 | 0.153 | 3.68 |
| *T* asymmetric | 1325.51 | $1.16\times10^5$ | 0.328 | $2.69\times10^4$ | $1.45\times10^3$ | 0.99931 | 0.34 | 0.927 | 2.99 |

The material parameters entering *C* carry uncertainty that should be distinguished from the simulated photonic quantities. For *AI1*, the adopted *DW= 0.14* is obtained from first-principles calculations,[23] while $\eta_{QE}\approx$ ≈ 0.20 is estimated from the measured 135-ns total lifetime[22] and calculated radiative lifetime.[23] The *T*-center quantum efficiency is likewise sensitive to defect configuration, with reported estimates ranging from approximately 0.18 for the hydrogenated center to near unity for its deuterated counterpart.[41] Because *C* scales linearly with DW× $\eta_{QE}$, these uncertainties change its absolute value but not the central high-cooperativity conclusion: the designs remain well above *C=1* under conservative material assumptions, with robustness to fabrication disorder examined explicitly in Section 2.4 and Section S6.

High cooperativity should, however, be distinguished from strong coupling. Cooperativity characterizes preferential emission into the cavity channel, whereas strong coupling requires coherent emitter–cavity exchange to exceed the relevant loss and decoherence rates. As a cavity-loss benchmark, we evaluate

$$\frac{2g}{\kappa} = \frac{Q}{Q_{\mathrm{crit}}} \tag{5}$$

where $g$ is the coherent emitter–cavity coupling rate, $\kappa$ is the cavity energy-decay rate, and $Q_{crit}$ is the quality factor for which $2g/\kappa=1$. Using $Q_{crit}$=1.6×10$^5$ for *Al1* and 3.4×10$^5$ for *T*, the asymmetric cavities give $2g/\kappa=1.8$ and 0.34, respectively, whereas the symmetric cavities reach approximately 20 and 6.0 (Fig. 3b). Thus, the symmetric designs and the asymmetric *Al1* design exceed the **c**avity-loss threshold $2g/\kappa=1$ in the idealized emitter model, while the asymmetric *T* design remains in the Purcell regime. Exceeding $2g/\kappa=1$ indicates that coherent emitter–cavity coupling can overcome cavity loss; however, experimental strong coupling additionally requires the coherent coupling rate to exceed the emitter decoherence rate, including homogeneous broadening and spectral diffusion. These emitter-dependent limitations are discussed in Section 2.6. Importantly, strong coupling is not required for the spin–photon interface considered here: the primary objective is high cooperativity, so that emission is preferentially coupled into the cavity mode, together with efficient extraction of those photons into the on-chip waveguide. For context, color-center nanobeam cavities in diamond have reached single-emitter Purcell factors of order ten,[42] whereas the silicon cavities designed here target the substantially larger enhancement required for a telecom-band spin–photon interface.

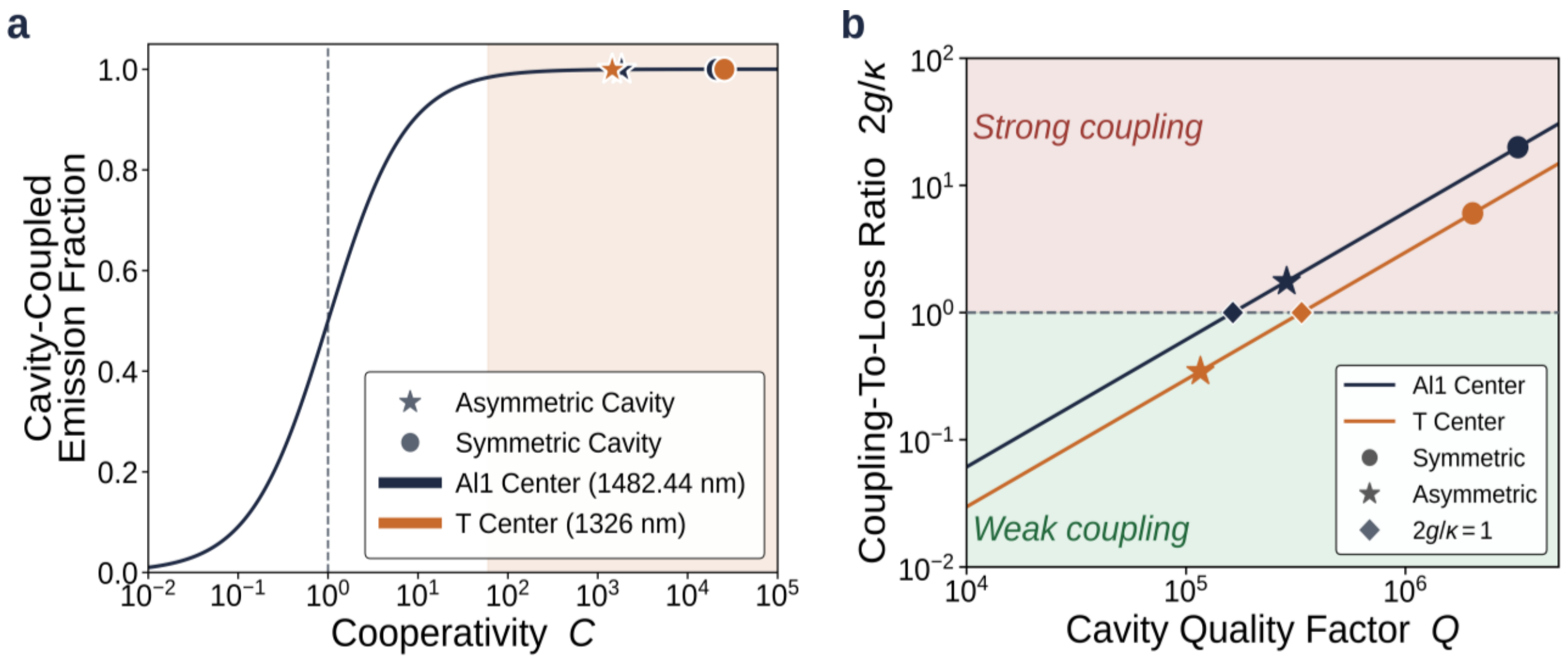

***Figure 3. Cavity-QED figures of merit for the T- and Al1-center nanobeam cavities.*** *(a) Cavity-coupled emission fraction $\beta_{cav} = C/(C+1)$ as a function of cooperativity $C = F_P \cdot DW \cdot \eta_{QE}$. All four designs lie in the saturated high-cooperativity regime, with $C \approx 2.1\times10^4$ to $2.6\times10^4$ for the symmetric cavities and $C \approx 1.4\times10^3$ to $1.8\times10^3$ for the asymmetric cavities. (b) Coupling-to-loss ratio $2g/\kappa = Q/Q_{crit}$ as a function of loaded cavity quality factor. The dashed line and diamonds mark the cavity-loss threshold $2g/\kappa = 1$. The region above the threshold, labeled strong coupling, marks the onset of strong coupling for an emitter at its radiative (homogeneous-linewidth) limit; the effect of the larger, spectral-diffusion-broadened linewidths measured for real devices is discussed in Section 2.6. The symmetric cavities and the asymmetric Al1-center cavity exceed this threshold, whereas the asymmetric T-center cavity remains below it. The emitter parameters used to calculate C and $Q_{crit}$ are summarized in Table 1 and Section S3.*

### *2.4. Robustness to nanometer-scale fabrication disorder*

High-Q nanobeam cavities are inherently sensitive to nanoscale geometric variations, making fabrication tolerance an important requirement for practical spin–photon interfaces. We therefore test all four optimized cavities against controlled perturbations of the hole pattern and re-evaluate their performance using well-resolved FDTD simulations at 18 cells per wavelength. Two complementary perturbation channels are applied to every design. The first is a uniform bias of all hole radii, swept over ±1, ±5 and ±10 nm to span the over- and under-etch range of realistic electron-beam lithography. The second is random per-hole disorder, in which the radius and the in-plane (x, y) position of every hole are drawn independently from a Gaussian of standard deviation $\sigma$ = 2 nm; ten independent realizations are generated per design, with seeds shared across the four cavities. This protocol probes both systematic etch bias and stochastic size-and-position disorder, and it isolates hole-pattern imperfection; sidewall-angle and device-layer-thickness variations are not included.

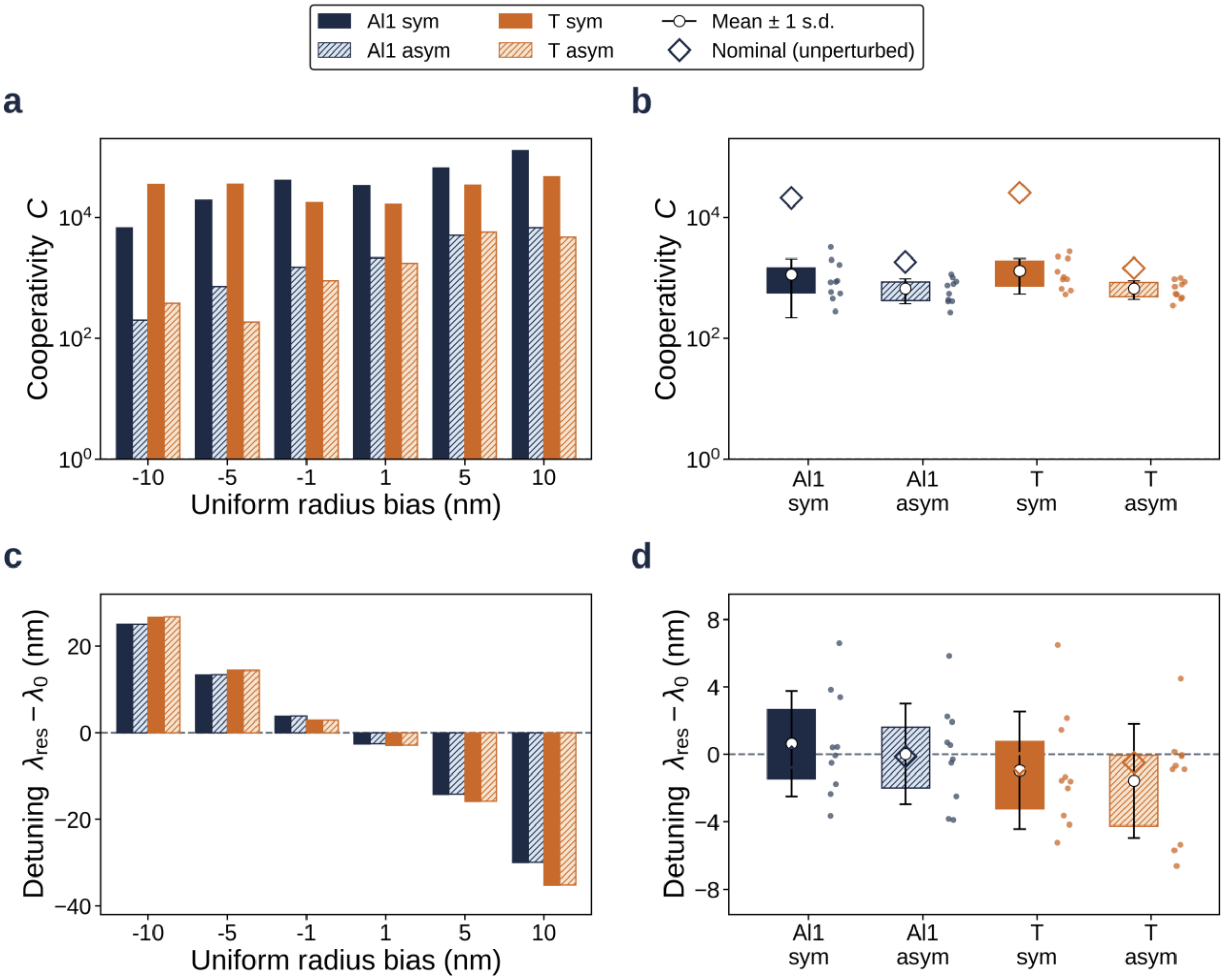

***Figure 4. Robustness of the inverse-designed cavities to hole-pattern perturbations. (a) Cooperativity C under a uniform bias of all hole radii of ±1, ±5 and ±10 nm (bar chart, a discrete perturbation), and (b) cooperativity for ten independent Gaussian per-hole realizations of radius and position at σ = 2 nm (box: interquartile range; open circle with whisker: mean ± 1 s.d.; points: individual realizations; open diamond: nominal unperturbed value). (c,d) Corresponding resonance detuning $\lambda_{res} - \lambda_0$ for the (c) uniform and (d) Gaussian perturbations. Cooperativity is computed with the adopted emitter quantum efficiencies ($\eta_{QE}$ = 0.234 for T and 0.20 for Al1). Across every perturbation the cavities remain deep in the high-cooperativity regime: the smallest value is C ≈ 187, for the asymmetric T-center cavity under a −5 nm uniform bias, not the strongest perturbation***

***applied, while the worst Gaussian σ = 2 nm realization gives C ≈ 271, both far above unity. Per-device ranges are listed in Table S3.***

Figure 4a,b shows the resulting cooperativity under the two perturbation channels. As expected for high-*Q* photonic-crystal cavities, geometric disorder produces substantial variations in the loaded quality factor and hence in *C*, particularly for the symmetric designs. Nevertheless, the cavity-QED performance remains well within the high-cooperativity regime across all four designs and all perturbations. The smallest cooperativity encountered is $C \approx 187$, for the asymmetric *T*-center cavity under a −5 nm uniform bias, which is not the strongest bias applied, while the ten Gaussian realizations at $\sigma = 2$ nm keep every design at a mean cooperativity of order $10^3$ with a worst single draw of $C \approx 271$. Because the nominal cooperativities are large, the corresponding cavity-coupled emission fraction $\beta_{cav} = C/(C+1)$ stays above 0.99 even in these worst cases, so the designs remain firmly in the $C \gg 1$ regime over the full range of perturbations tested. Because the objective is evaluated at the fixed color-center wavelength, perturbations that raise the bare cooperativity, such as the larger positive radius biases in Fig. 4a, do so by shifting the resonance away from the emitter line (Fig. 4c,d) rather than by improving the on-resonance performance; the non-monotonic response therefore reflects the detuning penalty built into the design objective and does not indicate that the nominal geometry is sub-optimal.

The perturbations also shift the cavity resonance (Fig. 4c,d): a ±1 nm uniform bias moves the resonance by roughly 3 to 4 nm, the larger ±5 and ±10 nm biases shift it by up to several tens of nanometers, and the $\sigma = 2$ nm per-hole disorder produces a spread of a few nanometers about the nominal line. Such spectral sensitivity is important because the linewidths of these high-*Q* cavities are much narrower than the resulting resonance shifts. Fabrication tolerance therefore does not eliminate the need for spectral alignment between the cavity and an individual color center; rather, it shows that the underlying cavity-enhancement performance survives nanometer-scale geometric variations. Practical implementation will require post-fabrication cavity or emitter tuning to restore spectral resonance.

Directional extraction in the asymmetric cavities is more sensitive to disorder than the cavity coupling itself. Under the uniform radius bias the right-directed fraction $\beta_R$ stays high (about 0.89 to 0.97 at ±1 nm), but the random per-hole disorder, which displaces the holes and thereby unbalances the two mirrors, reduces it substantially: at $\sigma = 2$ nm the mean $\beta_R$ falls to about 0.44 for the *T* asymmetric cavity and 0.35 for the *Al1* asymmetric cavity, with individual realizations as low as 0.18 (Table S3). The high nominal directionality near 0.9 is therefore a design value that presumes tight positional control or post-fabrication selection of well-formed devices, whereas the cavity-coupled emission fraction remains above 0.99 throughout. Fully building this positional tolerance into the geometry would require a robust-formulation inverse design that incorporates disorder directly into the optimization objective.[29,43] More broadly, these results show that the cavity enhancement is not confined to a single nominal geometry, while other fabrication imperfections, such as sidewall angle and device-layer thickness, remain to be evaluated.

### *2.5. Benchmarking against silicon color-center cavities*

Having established high cooperativity and robustness to nanometer-scale fabrication disorder, we next benchmark the inverse-designed cavities against reported silicon color-center cavity platforms (Fig. 5). To our knowledge, no cavity-integrated *Al1* center has yet been reported, making the present *Al1* structures the first cavity designs targeting this emitter. For the *T* center, several cavity architectures have been proposed and experimentally investigated. Because simulated and experimentally measured cavity metrics can differ substantially, Fig. 5 compares our simulated results with the corresponding design or simulated values reported for previous devices; experimentally measured performance is summarized separately in Section S7.

Figure 5a compares loaded quality factor and device footprint. The four inverse-designed cavities achieve $Q = 1.16\times10^5$- $3.25\times10^6$ within footprints below 4.5 μm$^2$. For comparison, the bus-coupled *T*-center symmetric cavity of Komza *et al.* has a design external $Q$ of order $10^5$,[44] while the asymmetric one-sided nanobeam of Islam *et al.* has a design $Q$=$1.17\times10^4$ at a comparable footprint.[45] The present designs therefore occupy a high-$Q$, compact-device region of the design space, with the symmetric cavities maximizing optical confinement and the asymmetric cavities retaining $Q > 10^5$ while providing a directional output channel.

The distinction becomes more relevant for an integrated spin–photon interface when Purcell enhancement is considered together with photon extraction (Fig. 5b). The asymmetric *Al1* and *T* cavities combine large Purcell factors, $F_P$ = $6.58\times10^4$ and $2.69\times10^4$ respectively, with right-directed waveguide fractions near 0.9. The published *T*-center devices provide useful reference points: Komza et al. report a guided out-coupling fraction of approximately 0.80 with a deduced cavity Purcell factor of approximately $1.5\times10^4$ [44] while Johnston et al. demonstrate one-sided *T*-center emission with a measured guided out-coupling fraction of approximately 0.36 and a simulated ZPL Purcell factor of approximately 470 [46] These comparisons illustrate the central objective of the asymmetric designs: retaining substantial cavity enhancement while coupling most of the cavity-mediated emission directly into a single integrated waveguide.

The complementary roles of the two cavity families are summarized in Fig. 5c. The symmetric cavities prioritize $Q$ and Purcell enhancement, whereas the asymmetric cavities trade part of that confinement for directional waveguide coupling, with only minor changes in mode volume. This comparison should nevertheless be interpreted at the design level. Experimentally realized silicon color-center cavities remain subject to fabrication disorder, absorption, emitter placement, spectral alignment, and other loss mechanisms not fully captured by idealized simulations. For example, related G-center experiments have demonstrated both symmetric photonic-crystal cavities[47] and asymmetric nanobeams with approximately 75% nanobeam-to-fiber coupling,[48] while experimentally realized emitter-loaded cavity quality factors remain substantially below the highest simulated values considered here. The present benchmark therefore identifies the performance targeted by the inverse-designed geometries rather than predicting the performance of fabricated devices directly.

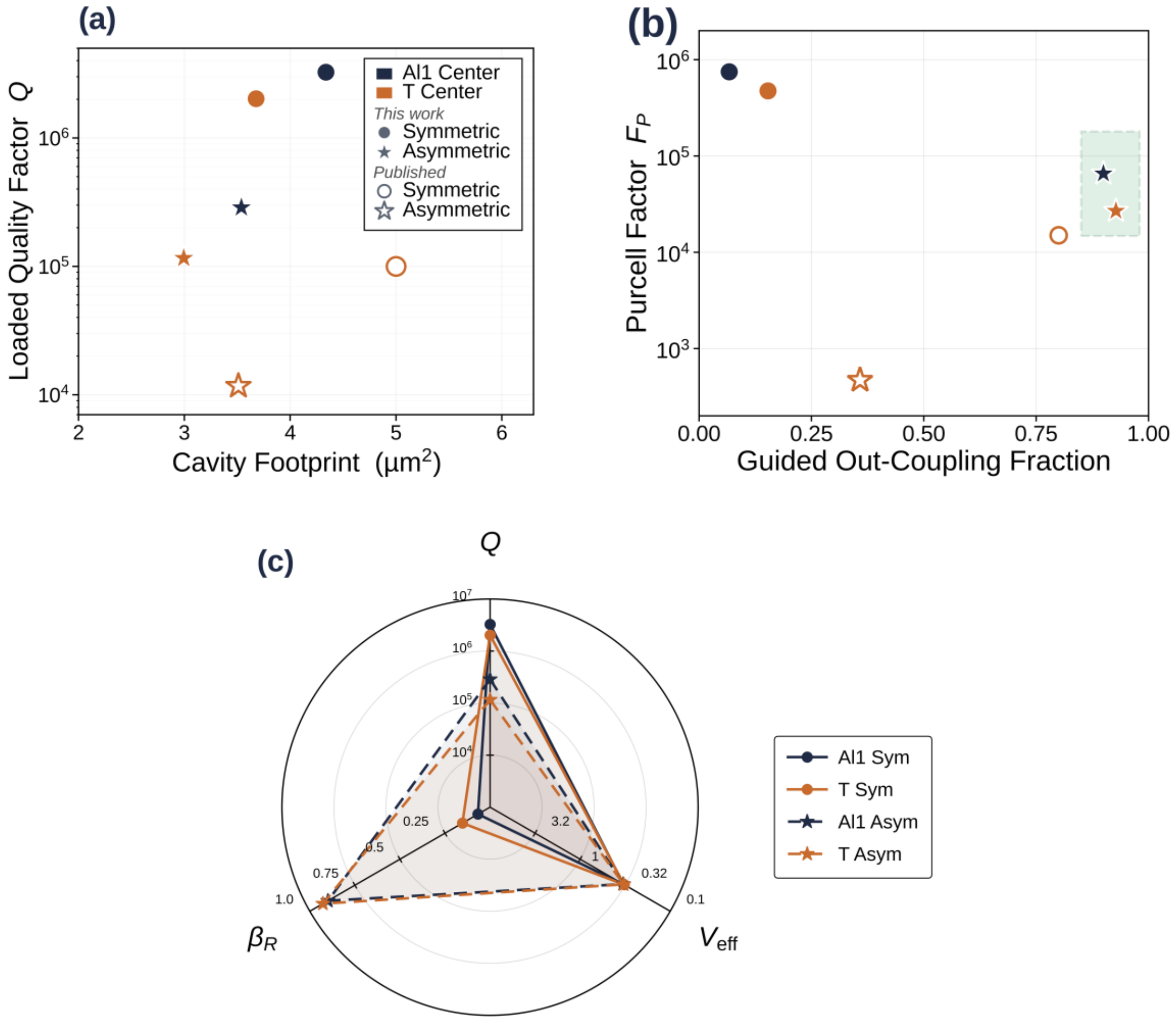

***Figure 5. Benchmarking of the inverse-designed cavities against reported silicon color-center cavity designs. Filled markers denote the simulated cavities of this work and open***

***markers previously reported devices; circles and stars denote symmetric and asymmetric architectures, respectively. (a) Loaded quality factor versus cavity footprint. Literature points correspond to reported design or simulated values for T-center cavities.[44,45] (b) Purcell factor versus guided out-coupling fraction, comparing the asymmetric cavities with reported T-center interfaces.[44,46] The asymmetric cavities of this work are highlighted (shaded region) because they are the operating point of interest for an integrated spin–photon interface, combining high Purcell enhancement with high guided out-coupling of about 0.9 at the nominal design. (c) Comparison of Q, $V_{eff}$, and right-directed fraction for the four cavities designed in this work. Literature values are compared on a design or simulation basis where possible; experimentally measured performance is summarized separately in Section S7. To our knowledge, no cavity design targeting the Al1 center has previously been reported.***

### *2.6. Realistic performance and path to experiment*

The simulated performance established above represents the intrinsic design capability of the cavities; experimentally realized devices will additionally be limited by fabrication disorder, material loss, and emitter dephasing. Three considerations therefore define the experimentally accessible operating regime. First, the simulated loaded quality factors, particularly those of the symmetric cavities, should be regarded as upper bounds. Silicon nanobeam cavities without color centers can reach intrinsic $Q>10^7$, whereas incorporation of implanted color centers can introduce absorption and other material losses related to the lattice-damage defects from ion implantation that limit the experimentally realized $Q$ to approximately $10^5$.[24] Geometric disorder is considered separately in Section 2.4, where hole-radius perturbations up to $\sigma$=2 nm alter the loaded $Q$ but preserve high cooperativity across all four designs. Notably, the same conclusion persists under a material-loss-limited $Q$: at $Q=10^5$, the mode volumes obtained here correspond to $C$~640–1260. Thus, the experimentally relevant high-cooperativity regime remains accessible even if the simulated ultra-high $Q$ is not fully retained. By contrast, the strong-coupling thresholds $Q_{crit} \approx 1.6\times10^5$ for *Al1* and $\approx 3.4\times10^5$ for *T* are above this conservative experimental level.

Second, the calculated cooperativities depend on the emitter properties through *DW* and $\eta_{QE}$. These quantities remain uncertain, particularly for the recently demonstrated *Al1* center, for which the material parameters are derived from a combination of theory and emerging experimental measurements. Section S3 therefore evaluates this uncertainty explicitly rather than treating the nominal values as exact. The high-cooperativity conclusion remains robust over the parameter ranges considered.

Third, the emitter linewidth imposes an additional constraint not captured by cavity $Q$ alone. The cavity-QED quantities in Table 1 are calculated using the cavity linewidth and lifetime-limited emitter parameters and therefore represent idealized upper bounds when spectral diffusion or pure dephasing is appreciable. For the symmetric cavities, $\kappa/2\pi$ = 62 MHz for *Al1* and 112 MHz for *T*, substantially narrower than the order-GHz laser-induced spectral diffusion reported for *T* centers in nanophotonic devices.[49] Correspondingly, the nominal symmetric-cavity cooperativities (~21,000–26,000) should not be interpreted as the effective cooperativities of a GHz-broadened emitter.

This distinction is particularly important for strong coupling. Beyond the cavity-loss condition $2g/\kappa > 1$ shown in Figure 3b, coherent strong coupling additionally requires the coherent coupling rate to exceed the emitter decoherence rate, that is $2g > \gamma^*$, where $\gamma^*$ includes homogeneous broadening and spectral diffusion; the coupling rate $g$ itself is set by the zero-phonon-line radiative rate and the effective mode volume, $g \propto 1/\sqrt{V_{\mathrm{eff}}}$, as defined in Section S2 (Equation (S12)). For the symmetric cavities, $2g/2\pi$ = 1240 MHz for *Al1* and 675 MHz for *T*; the emitter linewidths broadened by spectral diffusion, about 11 GHz for *Al1* and 1.1 to 4.6 GHz for *T* (Table S6), would therefore preclude strong coupling despite the cavity-only ratios $2g/\kappa > 1$. In this sense the present designs are cavity-limited rather than intrinsically weakly coupled: for an emitter at its radiative (homogeneous) linewidth, $2g$ exceeds both the cavity and the emitter decay rates, so strong coupling is in principle within reach[25], and the current barrier is the spectral-diffusion-broadened optical linewidth of these centers[22,49], a material-quality limitation, rather than the cavity design. A quantitative comparison of the symmetric-cavity coupling rates with the radiatively limited, homogeneous, and spectral-diffusion-broadened optical linewidths of both centers is given in Section S9. Under spectral-diffusion broadening, $2g/(\kappa + \gamma^*)$ falls below unity for all four cavities, which then operate in the weak-coupling (Purcell) regime (Table S7). The asymmetric cavities, with broader cavity linewidths $\kappa/2\pi \approx$ 0.7 to 2 GHz, are better matched to a spectrally broadened emitter. At a conservative absorption-limited $Q \sim 10^5$, the cavity linewidth is of order 2 GHz and the corresponding Purcell cooperativity remains 640–1260. High cooperativity and efficient directional extraction, rather than strong coupling, are therefore the experimentally relevant targets of the present designs. The required enhancement is also consistent with the broader experimental direction of silicon color-center photonics: hundred- to thousand-fold Purcell enhancement has been identified as a route toward overcoming emitter decoherence and approaching Fourier-limited optical interfaces,[24] while the initial *Al1* report identifies cavity integration as an important next step.[22] The present inverse-designed cavities provide a compact and fabrication-compatible platform for pursuing that regime experimentally.

## 3. Conclusion

We have developed a unified inverse-design framework for nanobeam cavities targeting telecom-band silicon spin–photon interfaces. Using a compact parametrization consisting only of the photonic-crystal lattice constants and a single fill factor, the same adjoint-optimization pipeline produces resonant cavities for both the *T* center at 1326 nm and the *Al1* center at 1482.44 nm. Two complementary cavity families emerge from this framework: symmetric cavities optimized for maximal Purcell enhancement and asymmetric cavities engineered for directional extraction, routing approximately 90% of the emitted power into a single on-chip waveguide.

The resulting designs reach the high-cooperativity regime required for efficient spin–photon interfacing. At the nominal simulated geometries, the four cavities provide cooperativities $C \sim 1.4\times10^3$ to $2.6\times10^4$ and cavity-coupled emission fraction $\beta_{cav}>0.999$, while the symmetric designs and the asymmetric *Al1* design additionally reach the cavity-defined strong-coupling regime. More importantly for experiment, high cooperativity remains robust against realistic hole-pattern disorder and persists at a conservative absorption-limited $Q$ about $10^5$, even though the ultra-high-$Q$ and strong-coupling predictions should be regarded as simulated upper bounds in the presence of material loss and emitter dephasing. The asymmetric cavities therefore provide a particularly practical operating point, combining high cooperativity with directional extraction of about 90% of the emitted power into an integrated photonic channel at the nominal design, subject to the positional-disorder sensitivity noted above.

The significance of the approach extends beyond the four devices demonstrated here. Because the inverse-design framework is not tied to a specific color center, the same optimization strategy can be retargeted to other silicon color centers by incorporating their emission wavelengths and emitter-specific optical properties into the design objective. More broadly, the methodology is not intrinsically restricted to silicon: by redefining the material platform, device geometry, target wavelength, and emitter properties, the same adjoint-optimization strategy can be extended to color-center cavity interfaces in other quantum photonic platforms, including diamond and silicon carbide. The framework therefore provides a general route toward cavity-enhanced spin–photon interfaces across multiple solid-state quantum-emitter platforms. For silicon, the compact geometry – a standard 220-nm silicon-on-insulator nanobeam described by only a small set of lattice parameters – provides a direct and fabrication-compatible path from inverse design to device realization. To our knowledge, these results provide the first cavity designs specifically targeting the *Al1* center and establish performance competitive with reported design values for *T*-center cavities at compact footprints.

The next step is experimental realization, including emitter incorporation, nanofabrication, spectral alignment, and characterization of material loss and emitter linewidth. Crucially, the central result does not depend on achieving the idealized ultra-high-*Q* limit: high cooperativity and efficient on-chip photon extraction remain accessible under experimentally realistic constraints. This combination establishes inverse-designed nanobeam cavities as a promising route toward integrated spin–photon interfaces for scalable quantum photonic circuits and networks, while providing a transferable design methodology for color-center quantum photonics beyond silicon.

**Appendix: Computational Section**

*Simulation:* All results are from three-dimensional finite-difference time-domain (FDTD) simulation of silicon nanobeams (refractive index $n_{Si}$ = 3.46, thickness 220 nm, width 500 nm) released as free-standing, air-clad membranes. The full three-dimensional domain is simulated without imposing symmetry boundary conditions, so that the shortened right mirror of the asymmetric cavities and the resulting directional out-coupling are captured without constraining the field to a prescribed symmetry. In every case the optimization targets the fundamental dielectric-band cavity mode of the nanobeam, whose in-plane electric field is polarized along the beam width and whose antinode is confined in the silicon at the taper center, coinciding with the intended color-center site. The emitter is modeled as a point dipole placed at the field antinode and aligned with the cavity-mode polarization (the y direction), representing the optimal orientation; a dipole misaligned by an angle θ reduces the Purcell factor and cooperativity by $\cos^2\theta$ without altering the cavity mode, and because silicon is optically isotropic the crystal orientation enters only through this angle. The loaded quality factor is extracted from the temporal ring-down of the cavity field: after the source is switched off, $\ln|E(t)|^2$ decays log-linearly and Q is read from the envelope slope (slope = $-\omega/Q$), rather than from a spectral fit (Section S1 and Figure S2). Effective mode volumes are computed from the simulated energy density and quoted in units of $(\lambda/n)^3$.

*Inverse design:* The geometry is optimized with the adjoint method. Each iteration runs one forward and one adjoint three-dimensional FDTD simulation in Tidy3D, whose automatic-differentiation plugin returns the gradient of the objective (Equation (1)) with respect to the design vector p = ($FF$, $a_{mir}$, $a_0$, ..., $a_4$), namely the fill factor, the mirror period and the five taper periods, and p is advanced by gradient ascent.[28,31] The figure of merit (Equation (1)), the update rule and the derivation of the adjoint gradient are given in Section S1. The two families per center follow from balanced or unbalanced mirrors. During the optimization loop the forward and adjoint simulations use a deliberately coarse spatial grid of 12 cells per wavelength, the optimization resolution, which keeps the per-iteration cost low; the quality factors, mode volumes and Purcell factors reported in Table 1 are recomputed for the converged geometries on a finer grid of 25 cells per wavelength, so that the quality factors monitored during optimization differ from these converged high-resolution values.

*Figures of merit:* The Purcell factor $F_P$, cooperativity $C = F_P \cdot DW \cdot \eta_{QE}$, cavity-coupled emission fraction $\beta_{cav} = C/(C+1)$, directional fraction $\beta_R$ and coupling-to-loss ratio $2g/\kappa = Q/Q_{crit}$ are defined and derived in Section S2. Debye-Waller factors and quantum efficiencies, with their sources and uncertainties, are listed in Section S3; the figures adopt DW = 0.23 and $\eta_{QE}$ = 0.234 for the T center and DW = 0.14 and $\eta_{QE}$ = 0.20 for the Al1 center. Full device metrics, the supporting fabrication-tolerance data and the benchmark basis are given in Sections S4 to S7, with the converged cavity geometries listed in Section S8 (Table S5).

## Supporting Information

Supporting Information is available from the Wiley Online Library or from the author.

## Acknowledgements

## Conflict of Interest

The authors declare no conflict of interest.

## Data Availability Statement

## Table of Contents Entry

Silicon has exactly two telecom color centers with an addressable ground-state spin, the T and Al1 centers, and each needs a cavity to become a spin-photon interface. A single adjoint inverse-design pipeline supplies that cavity layer for both, in a maximal-Purcell and a directional flavor, placing every design in the high-cooperativity regime a quantum network requires, robustly to fabrication.

## Supporting Information

This Supporting Information supplies the inverse-design method and the cavity-QED derivations that underpin the high-cooperativity claims of the main text, together with the material-parameter provenance, the full device metrics, the fabrication-tolerance study and the benchmark basis. All cavity figures of merit here are obtained from three-dimensional finite-difference time-domain (FDTD) simulation with adjoint inverse design. Material parameters (Debye-Waller factor DW, intrinsic quantum efficiency $\eta_{QE}$, radiative lifetime) are taken from the literature as stated. Simulated and measured quantities are labeled throughout. Cross-references of the form (Figure S1), (Table S1) and (Equation (S1)) are the anchors the main manuscript points to.

### S1. Inverse-design pipeline and simulation setup

Each cavity is a one-dimensional silicon photonic-crystal nanobeam: a single-mode silicon beam (refractive index $n_{Si}$ = 3.46, thickness t = 220 nm, width w = 500 nm, full beam length ≈ 12.1 μm) patterned with a row of air holes; this total length includes the unpatterned waveguide leads and is longer than the hole-patterned span that sets the footprint in Table S2a. The beam is released from the 220 nm silicon-on-insulator device layer to form a free-standing, fully air-clad nanobeam (a suspended membrane), so the dielectric environment is symmetric in the vertical direction. A central taper of $N_{taper}$ = 5 cells flanked by $N_{mir}$ = 7 mirror cells per side defines the localized mode.

**Parametrization.** The design variables are collected in a vector p = (FF, $a_{mir}$, $a_0$, ..., $a_4$): a single fill factor FF = r/a shared by all holes, the mirror period $a_{mir}$, and the five taper periods $a_0$ to $a_4$. Hole i is centered at the running sum of the periods and has radius $r_i$ = FF · $a_i$, so the positions and radii follow from p and nothing else is varied. For the asymmetric family the same vector is used, but the right mirror is shortened to three periods against seven on the left (NR = 3, NL = 7), deliberately weakening it to give the over-coupled output. The converged design vectors are listed in Table S5.

**Objective.** For a dipole at the beam-center field antinode, the optimizer maximizes a figure of merit O,

$$O = R\exp\left(-\frac{\delta^2}{2\sigma^2}\right) - \mu\,\delta^2 \qquad \text{(S1)}$$

whose reward R = $F_P$ for the symmetric cavity and R = $F_P$ · $\beta_R$ for the asymmetric one, so a single objective drives high Q/$V_{eff}$ and, where wanted, directional out-coupling, in the spirit of multi-objective inverse design of solid-state single-photon sources[1]. Here $\delta = (\lambda_{res} - \lambda_0)/\lambda_0$ is the fractional detuning from the target line $\lambda_0$; the Gaussian factor of width $\sigma = 10^{-3}$ rewards on-resonance geometries, and the quadratic penalty with adaptive coefficient $\mu = \kappa_q F_P$ ($\kappa_q = 10^6$) keeps the detuning term commensurate with the large Purcell reward. Together the two resonance terms lock the mode onto the color-center line while Q/$V_{eff}$ is maximized.

**Update.** The gradient of O with respect to p is obtained by the adjoint method (below), and p is advanced by a per-element, sign-based step,

$$p_i \leftarrow p_i + \eta\,|p_i|\,\mathrm{sign}\left(\frac{\partial O}{\partial p_i}\right) \qquad \text{(S2)}$$

with the learning rate η annealed from $10^{-3}$ to $10^{-5}$. This adjoint-based shape optimization[1-3] yields two families per center from the choice of mirror symmetry: a symmetric cavity (balanced mirrors, radiation spread transversely to the beam axis for off-chip collection) and an asymmetric cavity (unbalanced mirrors, in-plane directional out-coupling into a single waveguide).

**Adjoint gradient.** The objective depends on the design only through the permittivity distribution ε(r; p). Writing Maxwell's equations for the field E driven by the dipole source as a linear system A(ε) E = b, the field sensitivity follows from differentiating the constraint, $\partial E/\partial p_k = -A^{-1} (\partial A/\partial p_k) E$, and introducing the adjoint field $E_{adj}$ that solves the transposed system $A^T E_{adj} = (\partial O/\partial E)^T$. The design sensitivity then collapses to a single field overlap,

$$\frac{\partial O}{\partial p_k} = \mathrm{Re}\int E_{\mathrm{adj}} \cdot \frac{\partial \varepsilon}{\partial p_k} E \, \mathrm{d}V \tag{S3}$$

so the full gradient over all of p costs one forward and one adjoint FDTD simulation, independent of the number of design variables. Because ε depends on p through the hole edges (positions and radii $r_i = FF \cdot a_i$), the derivative $\partial\varepsilon/\partial p_k$ is localized on those boundaries; the automatic-differentiation plugin of the solver (Tidy3D) evaluates it and assembles the gradient $\partial O/\partial p$.

**FDTD.** All reported converged figures of merit come from high-resolution three-dimensional FDTD (Tidy3D) on an automatic nonuniform grid with at least 25 cells per wavelength, the free-standing beam extended 0.3 μm beyond its outermost holes and surrounded by 20-layer perfectly matched layers on all three axes with 0.5 μm axial and 0.8 μm transverse air padding; each simulation propagates for 100 ps with the field shutoff disabled. The mode is excited by a broadband Gaussian pulse (Tidy3D GaussianPulse) launched from the point dipole and centered on the target zero-phonon line (1326 nm for the T center and 1482.44 nm for the Al1 center), with a source spectral half-width of 5 nm; its temporal envelope has a characteristic width of about 0.1 ps, peaks near 0.5 ps and is switched off by about 1 ps, so the source is off well before the ring-down is read. The loaded quality factor is read from the temporal ring-down (below) and cross-checked against a resonance-fit estimator; per-face fluxes are sampled at 1000 frequencies over a ±3 nm window, and the effective mode volume from a field monitor whose time apodization starts at the source switch-off, so that the near field of the point dipole is excluded without any mask; the energy density is integrated with trapezoidal node weights on the nonuniform grid, each diagonal permittivity component being interpolated onto the nodes of the corresponding field component, and normalizing by the energy density at the emitter site rather than by its maximum changes the mode volume by less than 0.2%. The quality factors shown along the optimization trajectory (main-text Figure 1c,d, and Figure S1 for the symmetric family) are running values at the reduced optimization resolution of 12 cells per wavelength and differ from the converged high-resolution values reported here and in Table S2a; the fabrication-tolerance re-simulations (Section S6) use 18 cells per wavelength for tractability across the perturbation ensemble.

Figure S1 shows the convergence trajectories (resonance wavelength and Q versus iteration) for the symmetric cavities of both centers, whose resonances lock to the emitter zero-phonon lines at 1482.44 nm (Al1) and 1326 nm (T), complementing main-text Figure 1, which shows the asymmetric family as the representative case.

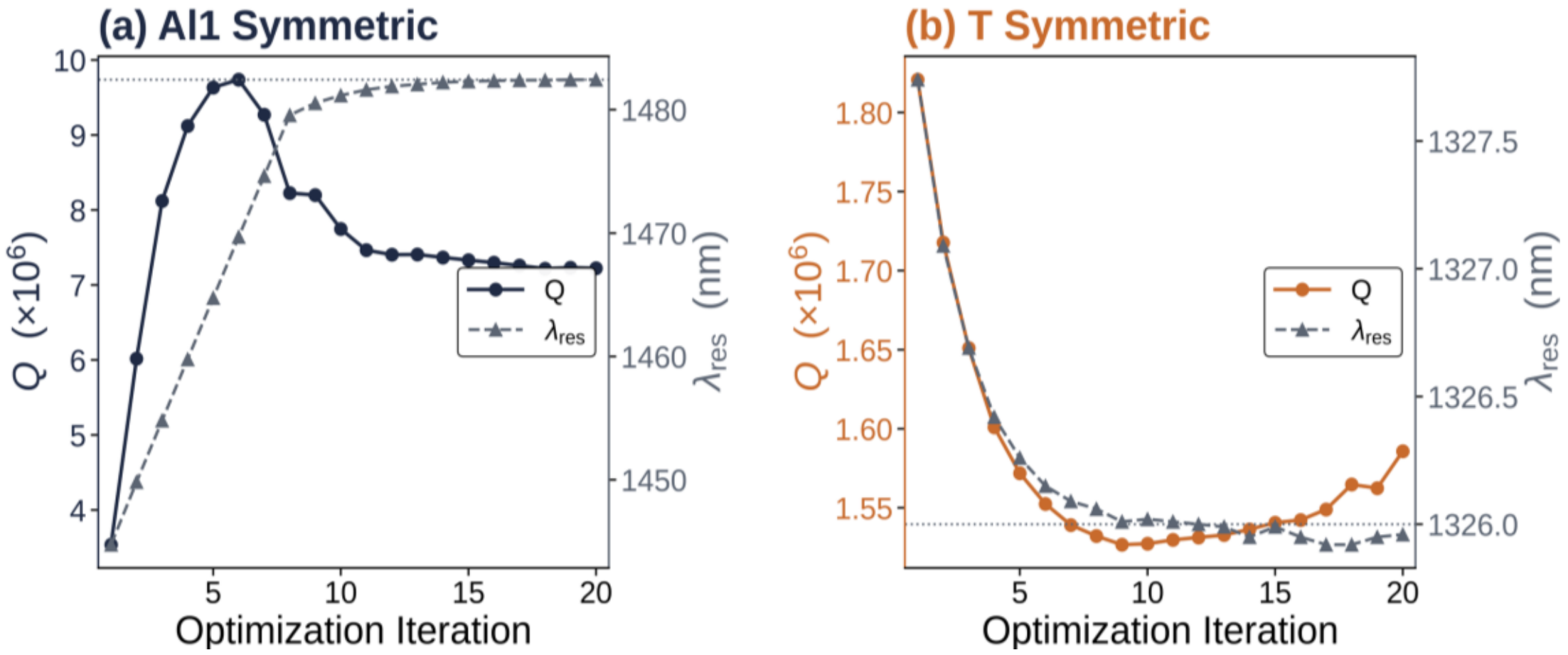

**Figure S1.** Convergence of the symmetric-cavity inverse design (all simulated). Loaded quality factor Q (left axis) and resonance wavelength $\lambda_{res}$ (right axis) versus optimization iteration for (a) the Al1 symmetric cavity, whose resonance converges to the emitter zero-phonon line, and (b) the T symmetric cavity. These are running values at the optimization resolution, distinct from the converged high-resolution figures of Table S2a.

The loaded quality factor is obtained from the temporal ring-down of the cavity field: once the source is switched off, the natural-log field intensity $\ln(|E(t)|^2)$ decays log-linearly and Q is read from the envelope slope (slope = $-\omega/Q$). Figure S2 shows the ring-down of the four cavities; the very shallow slopes of the symmetric cavities are the signature of their ultra-high loaded Q (Table S2a).

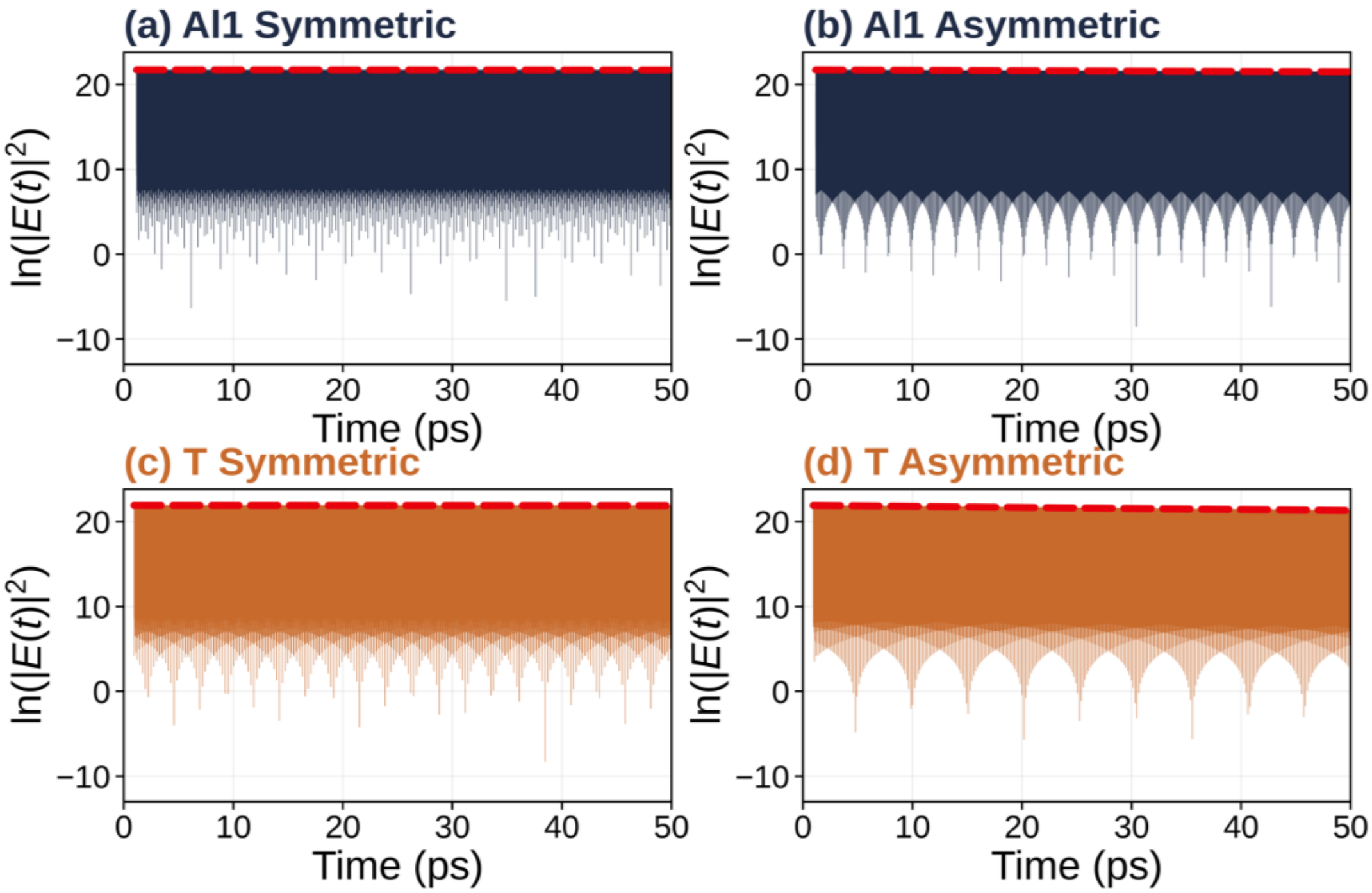

**Figure S2.** Cavity temporal ring-down and quality-factor extraction (all simulated). Natural-log field intensity $\ln(|E(t)|^2)$ versus time after the source is switched off, over the first 50 ps: the filled band is the full oscillating field intensity, its upper edge is the decay envelope, and the dashed

red line is the log-linear fit whose slope gives Q. The loaded quality factor follows from the envelope slope = −ω/Q; the extracted quality factors are listed in Table S2a. The very shallow slope of the symmetric cavities is the signature of their ultra-high Q.

## S2. Cavity-QED figures of merit: definitions and derivations

For a dipole at the cavity field maximum, on resonance, the Purcell factor is

$$F_P = \frac{3}{4\pi^2}\left(\frac{\lambda}{n}\right)^3 \frac{Q}{V_{\mathrm{eff}}} \tag{S4}$$

with the effective mode volume $V_{eff}$ expressed in units of $(\lambda/n)^3$ and given by Veff = $\int \varepsilon|E|^2\, dV / \max(\varepsilon|E|^2)$, evaluated on the converged FDTD field; the open-cavity (quasinormal-mode) correction is negligible at these quality factors. The purely photonic coupling fraction is

$$\beta_{\mathrm{temp}} = \frac{F_P}{F_P+1} \tag{S5}$$

Folding in the Debye-Waller factor DW and the intrinsic quantum efficiency $\eta_{QE}$ gives the emitter cooperativity and the cavity-coupled emission fraction

$$C = F_P \cdot \mathrm{DW} \cdot \eta_{\mathrm{QE}} \tag{S6}$$

$$\beta_{\mathrm{cav}} = \frac{C}{C+1} \tag{S7}$$

where $\beta_{cav}$ is the fraction of the total emitter decay (zero-phonon line, phonon sideband and non-radiative channels) funneled into the cavity mode through the zero-phonon transition. This factorization of the emitter cooperativity into the photonic Purcell factor and the zero-phonon radiative fraction DW·$\eta_{QE}$ is the standard cavity-QED construction for color-center interfaces. The guided directional fraction is

$$\beta_R = \frac{P_{\mathrm{right}}}{P_{\mathrm{total}}} \tag{S8}$$

the fraction of the total radiated power emitted through the right-hand output face. Within the standard cavity-QED framework[4], the coupling-to-loss ratio is

$$\frac{2g}{\kappa} = \frac{Q}{Q_{\mathrm{crit}}} \tag{S9}$$

$$Q_{\mathrm{crit}} = \sqrt{\frac{4\pi^2 V_{\mathrm{eff}}\,\omega}{3\,\gamma_{\mathrm{ZPL}}}} \tag{S10}$$

with ω = 2πc/λ, and the zero-phonon-line radiative rate

$$\gamma_{\mathrm{ZPL}} = \frac{\mathrm{DW}\cdot\eta_{\mathrm{QE}}}{\tau_{\mathrm{total}}} = \frac{\mathrm{DW}}{\tau_{\mathrm{rad}}} \tag{S11}$$

Numerically $\gamma_{ZPL}$(Al1) = 2.0×$10^5$ $s^{-1}$ and $\gamma_{ZPL}$(T) = 5.4×$10^4$ $s^{-1}$, giving $Q_{crit}$(Al1) = 1.6×$10^5$ and $Q_{crit}$(T) = 3.4×$10^5$.

**Cooperativity from the cavity-QED definition.** Equation (S4) for the Purcell factor and the coupling-to-loss ratio are the same physics seen two ways. In cavity QED the single-emitter cooperativity is C = $4g^2/(\kappa\gamma)$, with κ = ω/Q the cavity energy-decay rate, γ = $1/\tau_{total}$ the total emitter decay rate, and g the emitter-cavity coupling, set for a dipole at the field antinode by the zero-phonon-line radiative rate,

$$g^2 = \frac{3\,\omega\,\gamma_{\mathrm{ZPL}}}{16\pi^2 V_{\mathrm{eff}}} \tag{S12}$$

with $V_{eff}$ in units of $(\lambda/n)^3$. Using $\kappa = \omega/Q$ and $\gamma_{ZPL} = DW \cdot \eta_{QE} \cdot \gamma$ from the rate above,

$$C = \frac{4g^2}{\kappa\gamma} = \frac{\gamma_{\mathrm{ZPL}}}{\gamma}\frac{3}{4\pi^2}\frac{Q}{V_{\mathrm{eff}}} = F_P \cdot \mathrm{DW} \cdot \eta_{\mathrm{QE}} \tag{S13}$$

so the working form $C = F_P \cdot DW \cdot \eta_{QE}$ is exactly the cavity-QED cooperativity, and the cooperativity panel (Figure 3a) and the coupling-ratio panel (Figure 3b) both follow from the same g, κ and γ. The dipole-to-rate link follows from the Wigner-Weisskopf relation as applied in[5]. For an ideal two-level emitter ($DW = \eta_{QE} = 1$) this reduces to $C = F_P$. We adopt the convention $C = 4g^2/(\kappa\gamma)$, with κ and γ the full energy-decay rates; other conventions (for example $C = g^2/(\kappa\gamma)$) differ only by a fixed numerical factor and change none of the conclusions here, which depend on $C \gg 1$ and on the convention-independent coupled-emission fraction $\beta_{cav} = C/(C+1)$.

## S3. Material parameters and their provenance

Table S1 lists the Debye-Waller factor, quantum efficiency and lifetimes used for the two centers.

**Table S1.** Debye-Waller factor, quantum efficiency and lifetimes used for the two centers.

| Center | ZPL (nm) | DW | $\eta_{QE}$ (adopted) | $\eta_{QE}$ range | $\tau_{rad}$ | $\gamma_{ZPL}$ ($s^{-1}$) | Sources |
|---|---|---|---|---|---|---|---|
| T | 1326 | 0.23 | 0.234 | 0.181 to 0.984 | ≈ 4.3 μs | $5.4\times10^4$ | DW: Bergeron[6]; $\eta_{QE}$: Johnston[7] (0.234), Kazemi[8] (0.181 hydrogenated, 0.984 deuterated) |
| Al1 | 1482.44 | 0.14 (DFT DWF = 0.136) | 0.20 | n/a (single value) | 0.70 μs | $2.0\times10^5$ | DWF, $\tau_{rad}$, TDM = 1.81 D: Xiong[5] (DFT); $\tau_{total}$ = 135 ns: Crosta[9]; $\eta_{QE} = \tau_{total}/\tau_{rad}$ = 135/700 ≈ 0.20 |

For the Al1 center neither DW nor $\eta_{QE}$ is measured directly. The Debye-Waller factor is the DFT value (DWF = 0.136) of Xiong et al.[5]; the quantum efficiency $\eta_{QE} \approx 0.20$ is reconstructed as the ratio of the measured single-emitter total lifetime $\tau_{total}$ = 135 ns (Crosta et al.[9]) to the DFT radiative lifetime $\tau_{rad}$ = 0.70 μs (Xiong et al.[5]), and therefore combines theory and experiment. The two routes are consistent: $\gamma_{ZPL} = DW \cdot \eta_{QE}/\tau_{total} = DW/\tau_{rad} = 0.14 / 0.70\ \mu s = 2.0\times10^5\ s^{-1}$.

**$\eta_{QE}$(T) robustness.** The figures adopt $\eta_{QE}(T) = 0.234$, the lower bound of Johnston et al.[7] Kazemi et al.[8] place the hydrogenated center at $\eta_{QE} \approx 0.181$ and the deuterated center near unity (0.984). Because $C = F_P \cdot DW \cdot \eta_{QE}$, cooperativity stays far above one across this whole range: the smallest device (T asymmetric, $F_P = 2.69\times10^4$, DW = 0.23) gives C = 1120, 1447 and 6087 at $\eta_{QE}$ = 0.181, 0.234 and 0.984 respectively; the largest (T symmetric, $F_P = 4.74\times10^5$) gives $C = 1.97\times10^4$, $2.55\times10^4$ and $1.07\times10^5$. Every device satisfies $C \gg 1$ (minimum C = 1120) irrespective of the $\eta_{QE}$ adopted, so the conclusions of Figure 3 do not depend on this choice. DW and $\eta_{QE}$ are kept as separate factors; their product is the zero-phonon-line radiative fraction.

## S4. Full device figures of merit

Table S2a collects the high-resolution simulated figures of merit of the four cavities (all values simulated). C and $\beta_{cav}$ use DW and $\eta_{QE}$ from Table S1; 2g/κ uses $Q_{crit}$ from Section S2. A radiation-channel decomposition of the loaded quality factor into its waveguide-coupling and scattering contributions, together with the individual per-face quality factors, is given in Table S2b.

**Table S2a.** High-resolution simulated figures of merit of the four cavities (all values simulated).

| Device | $\lambda_{res}$ (nm) | Q (loaded) | $V_{eff}$ $((\lambda/n)^3)$ | $F_P$ | C | $\beta_{cav}$ | 2g/κ | $\beta_R$ | Footprint ($\mu m^2$) |
|---|---|---|---|---|---|---|---|---|---|
| Al1 symmetric | 1482.26 | $3.25\times10^6$ | 0.330 | $7.49\times10^5$ | $2.10\times10^4$ | 1.000 | 20 | 0.067 | 4.34 |
| Al1 asymmetric | 1482.32 | $2.87\times10^5$ | 0.332 | $6.58\times10^4$ | $1.84\times10^3$ | 0.999 | 1.8 | 0.899 | 3.54 |
| T symmetric | 1325.45 | $2.02\times10^6$ | 0.324 | $4.74\times10^5$ | $2.55\times10^4$ | 1.000 | 6.0 | 0.153 | 3.68 |
| T asymmetric | 1325.51 | $1.16\times10^5$ | 0.328 | $2.69\times10^4$ | $1.45\times10^3$ | 0.999 | 0.34 | 0.927 | 2.99 |

**Table S2b.** Radiation-channel decomposition of the loaded quality factor (all simulated, values in units of $10^6$). The two waveguide faces (right, left) group into Qwg and the four transverse faces (top, bottom, front, back) into Qscat, with 1/Qtotal = 1/Qwg + 1/Qscat. From the symmetric to the asymmetric design of each center Qwg falls by a factor of 53 (T) and 77 (Al1) while Qscat is essentially unchanged (within a factor of 1.2 to 1.6), so the reduced loaded Q of the asymmetric cavities is the intended waveguide coupling rather than added loss.

| Device | $Q_{total}$ | $Q_{wg}$ | $Q_{scat}$ | $Q_{right}$ | $Q_{left}$ | $Q_{top}$ | $Q_{bottom}$ | $Q_{front}$ | $Q_{back}$ |
|---|---|---|---|---|---|---|---|---|---|
| T symmetric | 2.02 | 6.61 | 2.92 | 13.2 | 13.2 | 7.81 | 7.81 | 25.5 | 21.1 |
| T asymmetric | 0.116 | 0.124 | 1.83 | 0.125 | 12.4 | 4.93 | 4.93 | 15.2 | 13.4 |
| Al1 symmetric | 3.25 | 24.3 | 3.76 | 48.6 | 48.6 | 14.3 | 14.3 | 13.2 | 19.6 |
| Al1 asymmetric | 0.287 | 0.317 | 3.03 | 0.320 | 47.9 | 9.54 | 9.53 | 14.3 | 20.0 |

Footprint is the hole-pattern span times the beam width (0.5 μm). Intrinsic (unloaded) quality factors exceed the loaded values by ~10× for the over-coupled asymmetric cavities; the symmetric cavities are under-coupled with loaded Q ≈ intrinsic Q (for example $Q_{int} \approx 3.5\times10^6$ and $2.9\times10^6$ for the Al1 symmetric and asymmetric cavities).

The effective mode volume of Equation (S5) is set by the tightly confined field at the central taper, which is common to the symmetric and asymmetric designs, and not by the terminal mirrors that control the quality factor. Weakening the right mirror does let the asymmetric-cavity mode extend further toward the output, with the in-plane field amplitude at the right edge of the field monitor exceeding that at the left by more than an order of magnitude, but this leaked field stays below $10^{-4}$ of the peak energy density and contributes less than 0.1% of the integral. Figure S3 confirms that $V_{eff}$ is therefore well-defined and box-independent: for all four cavities it rises as the integration box captures the confined core and then saturates onto a clean plateau for boxes longer than 3 μm, and the symmetric and asymmetric value of each center agree to within 1.2% (T: 0.324 versus 0.328; Al1: 0.330 versus 0.332), the level at which $V_{eff}$ is fixed by the shared central taper. Confinement and mirror coupling are thus largely independent design knobs: the resulting change in the coherent coupling rate g  $1/\sqrt{V_{eff}}$ is below 0.6%, leaving all cooperativities unchanged at the level relevant to the C 1 conclusion.

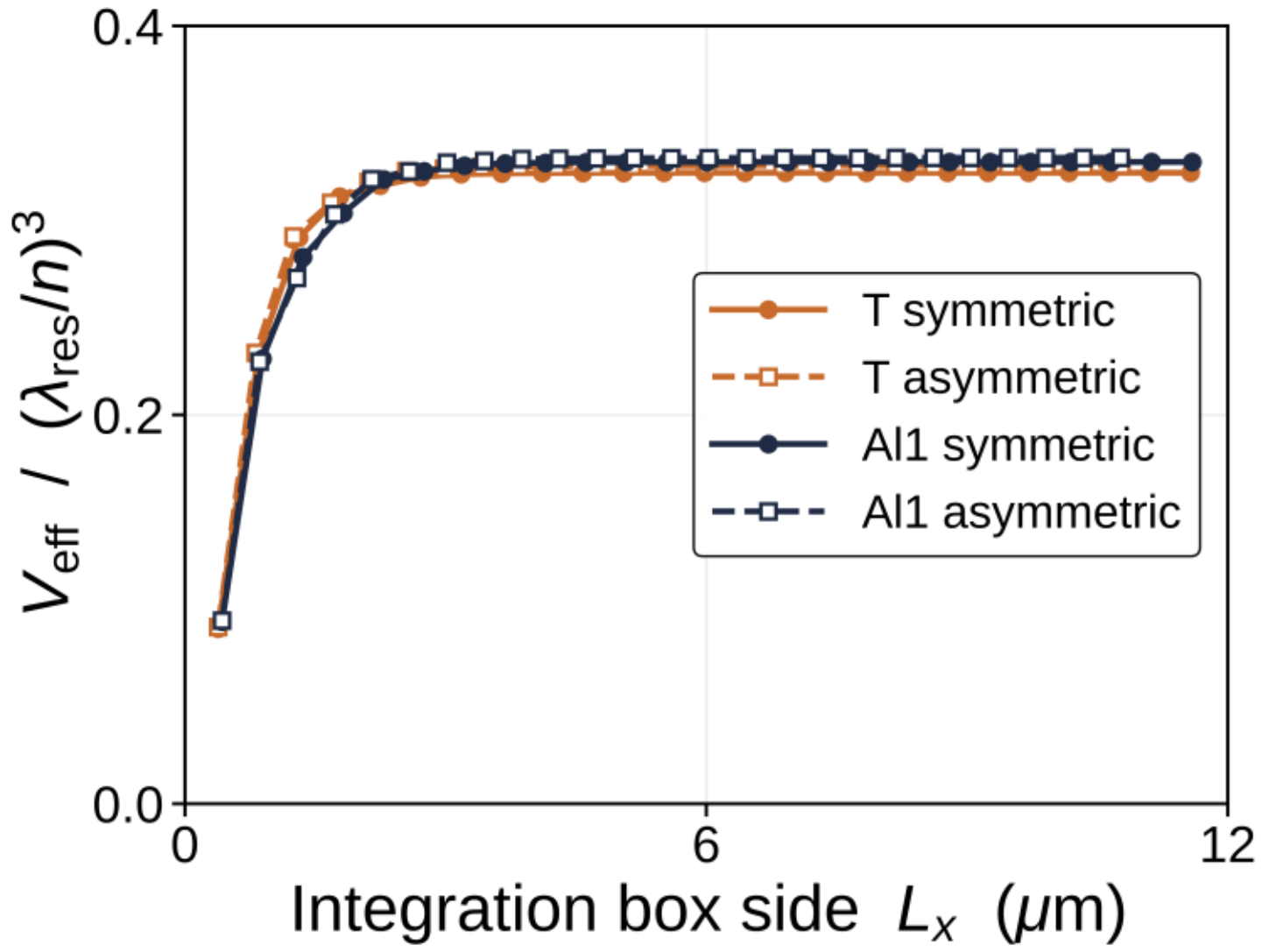

**Figure S3.** Convergence of the effective mode volume with the integration-box size (all simulated). Effective mode volume $V_{eff}$ (Equation (S5)), in units of $(\lambda_{res}/n)^3$, versus the side $L_x$ of the real-space integration box (centered on the cavity and extended along the beam axis) over which the energy-density integral is evaluated, for the symmetric (solid, filled markers) and asymmetric (dashed, open markers) designs of the T center (orange) and Al1 center (navy). All four curves saturate for $L_x$ 3 μm; the plateau values coincide with the $V_{eff}$ reported in Table S2a.

**S5. Flux-decomposition visualization (Figure 2c,d)**

Panels 2c and 2d are not a simulated three-dimensional field; they are a visualization that shows, at a glance, through which faces of the beam each cavity radiates. Starting from the FDTD-simulated in-plane field $Re(E_y)$ in the xy plane at z = 0, each face of the nanobeam is extruded outward by an amount proportional to the fraction of total emitted power radiated through that face (the per-face flux fractions $P_{top}$, $P_{bottom}$, $P_{front}$, $P_{back}$, $P_{left}$, $P_{right}$), with $Re(E_y)$ mapped onto color (color scale shared by the two panels). The symmetric cavity shows balanced, largely transverse leakage (top and bottom faces essentially equal, consistent with the suspended, vertically symmetric geometry); the asymmetric cavity concentrates the flux into the right-hand output face ($\beta_R = P_{right}/P_{total}$). Per-face fractions of the converged geometries are listed in the flux-decomposition data files.

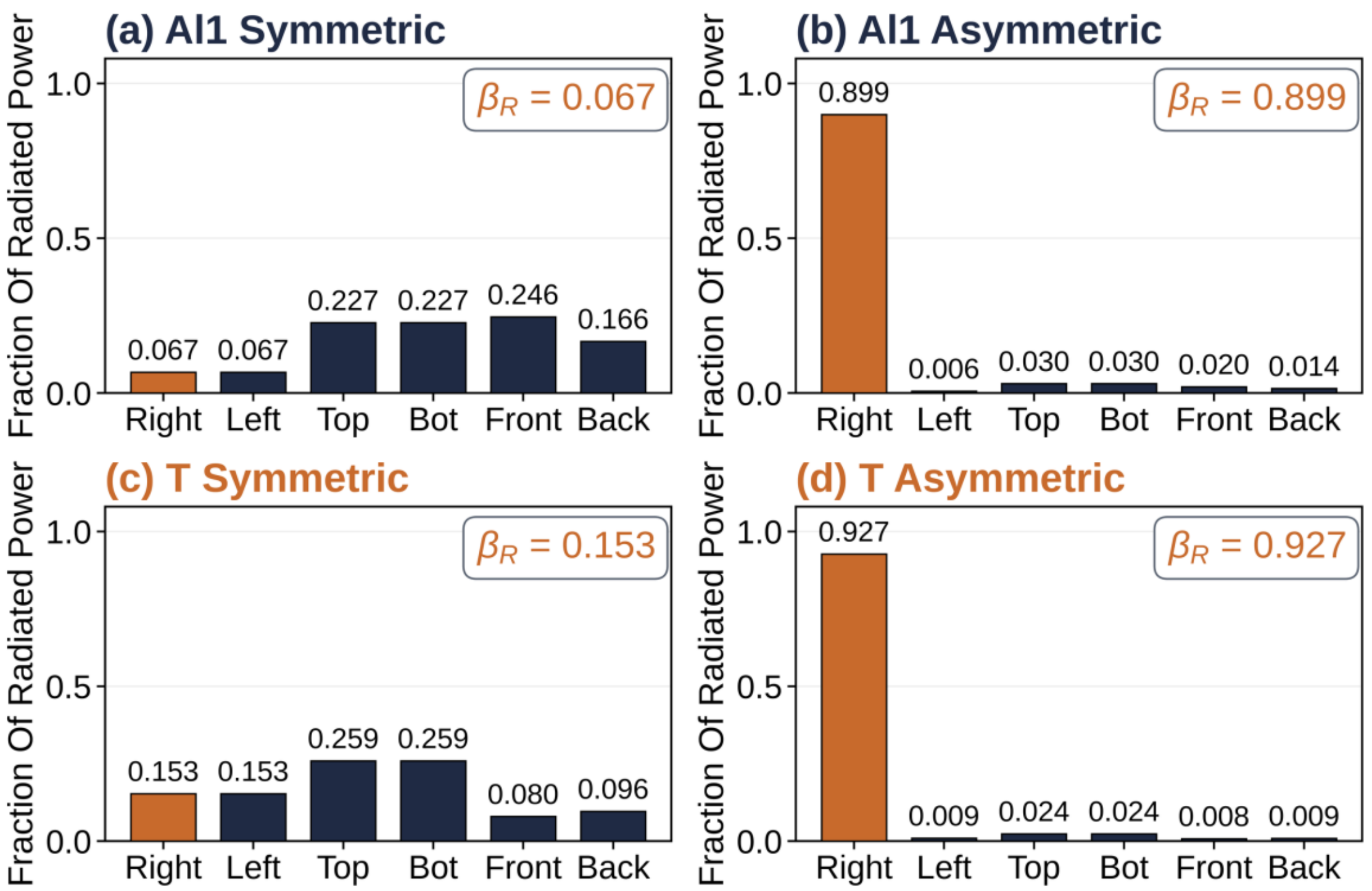

**Figure S4.** Per-face flux decomposition of the four cavities (all simulated). Fraction of the total radiated power exiting each of the six faces (Right, Left, Top, Bottom, Front, Back); the right output face is highlighted and its fraction $\beta_R$ is quoted in each panel. The symmetric cavities (a, c) radiate mostly out of plane with equal top and bottom faces, while the asymmetric cavities (b, d) send most of the power to the right face. Same data as the visualization in main-text Figure 2c,d.

## S6. Fabrication-tolerance study

Robustness is assessed by re-simulating all four optimized cavities on a well-resolved grid of 18 cells per wavelength under two perturbation channels. The first is a uniform bias of every hole radius, swept over ±1, ±5 and ±10 nm to span the over- and under-etch range of realistic electron-beam lithography. The second is random per-hole disorder in which the radius and the in-plane (x, y) position of every hole are drawn independently from a Gaussian of standard deviation $\sigma$ = 2 nm, with ten independent realizations per design and the random seeds shared across the four cavities. This gives sixteen perturbed simulations per cavity. Cooperativity stays well above unity throughout: the smallest value is C ≈ 187 (asymmetric T-center cavity, −5 nm uniform bias) and the worst Gaussian draw is C ≈ 271, so βcav remains above 0.99. The directional fraction βR of the asymmetric cavities is more fragile: it stays near 0.9 to 0.97 under ±1 nm bias but, because the random displacements unbalance the two mirrors, its mean falls to 0.44 (T) and 0.35 (Al1) at $\sigma$ = 2 nm, with individual realizations as low as 0.18.

**Table S3.** Fabrication-tolerance figures of merit of the four cavities under the sixteen perturbations of Section 2.4 (all simulated). C uses the adopted quantum efficiencies (T: 0.234, Al1: 0.20). The uniform range spans the ±1/±5/±10 nm radius biases; the $\sigma$ = 2 nm columns give the mean and worst of the ten Gaussian per-hole realizations of radius and position. βR is meaningful only for the asymmetric designs.

| Device | C nom | C (uniform ±1..±10 nm) | C at σ=2 nm (mean; worst) | βR nom | βR at σ=2 nm (mean; min) | λres range (nm) |
|---|---|---|---|---|---|---|
| T symmetric | 25524 | 16317 to 47261 | 1317; 533 | 0.15 | 0.03; 0.02 | 1291 to 1353 |
| T asymmetric | 1447 | 187 to 5767 | 667; 347 | 0.93 | 0.44; 0.21 | 1291 to 1353 |
| Al1 symmetric | 20984 | 6697 to 126181 | 1142; 280 | 0.07 | 0.03; 0.01 | 1452 to 1508 |
| Al1 asymmetric | 1841 | 201 to 6776 | 667; 271 | 0.90 | 0.35; 0.18 | 1452 to 1508 |

**S7. Benchmark data and comparison basis**

The published silicon T-center nanobeam cavities compared in main-text Figure 5 are simulated and measured on different bases; the points are plotted on a simulated or design basis so that they meet our simulated cavities on the same footing, with the measured values given here for context.

Table S4 lists each published point in Figure 5, the exact quantity plotted, and its basis. The "this work" points (Table S2a) are all simulated.

**Table S4.** Published points in Figure 5, exact quantity plotted and its basis.

| Reference | Panel(s) | Cavity | Quantity plotted (value) | Basis | Measured counterpart |
|---|---|---|---|---|---|
| Komza et al.[10] | 5a, 5b | symmetric, bus-coupled (T) | external Q ≈ $1×10^5$ (5a); $F_P$ ≈ $1.5×10^4$ (5b) at guided efficiency 0.80 | design Q; $F_P$ deduced from design Q ($C_0 \cdot Q_e/V$); guided efficiency measured | measured intrinsic Q ≈ $3.5×10^4$, measured Purcell $F_P$ ≥ 61, fiber efficiency ≈ 40% |
| Islam et al.[11] | 5a | asymmetric one-sided (T) | design loaded Q = $1.17×10^4$ at footprint ≈ 3.5 μm$^2$ | design Q (simulated); footprint is our estimate from the published geometry | measured loaded Q = $6.6×10^3$, measured Purcell F > 18 |
| Johnston et al.[7] | 5b | one-sided (T) | $F_P$ = 470 at guided out-coupling 0.358 | $F_P$ simulated ($P_{sim,ZPL}$, ideal placement); out-coupling measured | measured loaded Q = $4.3×10^4$, measured Purcell ≥ 25.6 |

**Reality check.** Empty silicon nanobeam cavities reach intrinsic Q > $10^7$, whereas emitter-loaded devices are limited to Q ≲ $10^5$ in practice by implantation-related absorption[12]. Even at an absorption-limited loaded Q = $10^5$, our mode volumes give C ≈ 640 (Al1) to ≈ 1260 (T): the cavities remain far inside the high-cooperativity regime, that is, the C ≫ 1 conclusion survives the realistic Q ceiling. A distinct, excitation-induced limitation is the roughly GHz laser-induced spectral diffusion of T centers in nanophotonic devices[13]; the centers are otherwise dark-stable on millisecond timescales.

### S8. Converged cavity geometries

The converged design vector for each of the four cavities is given in Table S5; the full hole pattern (positions and radii ri = FF ai) follows from it, and each geometry maps directly onto a silicon-on-insulator fabrication mask.

**Table S5.** Converged design parameters of the four inverse-designed cavities (all simulated). FF is the shared fill factor r/a, amir the mirror period, and a0 to a4 the five taper periods; each hole radius is ri = FF·ai. Mirror-hole counts per side are NL/NR (NL = NR = 7 symmetric; NL = 7, NR = 3 asymmetric, the shortened right mirror giving the directional out-coupling), and the taper count is Ntaper = 5. Periods in nanometers.

| Device | NL/NR | FF | amir (nm) | a0 (nm) | a1 (nm) | a2 (nm) | a3 (nm) | a4 (nm) |
|---|---|---|---|---|---|---|---|---|
| AI1 symmetric | 7/7 | 0.270 | 399.83 | 295.98 | 321.94 | 347.91 | 373.87 | 399.83 |
| AI1 asymmetric | 7/3 | 0.270 | 399.85 | 295.99 | 321.96 | 347.92 | 373.89 | 399.85 |
| T symmetric | 7/7 | 0.281 | 341.48 | 241.59 | 266.46 | 291.55 | 316.52 | 341.31 |
| T asymmetric | 7/3 | 0.281 | 341.40 | 241.58 | 266.51 | 291.51 | 316.49 | 341.35 |

### S9. Emitter optical decoherence and the strong-coupling regime

The main text establishes that all four cavities operate deep in the high-cooperativity regime and that the symmetric cavities and the asymmetric AI1 cavity additionally satisfy the cavity-loss condition $2g/\kappa > 1$ (Fig. 3b). Coherent strong coupling requires, in addition, that the coherent coupling rate exceed the emitter optical decoherence rate, $2g > \gamma^*$, where $\gamma^*$ is the full optical linewidth of the emitter under operating conditions and combines the homogeneous (lifetime- and phonon-limited) linewidth with inhomogeneous spectral diffusion. Because g is fixed by the cavity design, through the mode volume, whereas $\gamma^*$ is set by the emitter and its charge and thermal environment, the two conditions probe different limitations. Below we compare the symmetric-cavity coupling rates ($2g/2\pi$ = 1240 MHz for the AI1 center and 675 MHz for the T center) against the optical linewidths reported for the two centers across the relevant regimes.

We distinguish four linewidth regimes. (i) The radiative (transform) limit $\Delta\nu = 1/(2\pi\tau)$, set by the excited-state lifetime $\tau$: 0.17 MHz for the T center ($\tau$ = 0.94 μs) and 1.2 MHz for the AI1 center ($\tau$ = 135 ns). (ii) The homogeneous linewidth at the typical operating temperatures of 3 to 4 K; for the T center this is dominated by thermally activated mixing between the $TX_0$ and $TX_1$ bound-exciton states (splitting 1.76 meV)[6] and is therefore strongly temperature dependent, whereas for the AI1 center it has been measured directly. (iii) The optical linewidth of single emitters, broadened by spectral diffusion. (iv) The narrowest linewidth demonstrated after active charge stabilization. The corresponding coupling ratios are collected in Table S6.

**Table S6.** Coherent coupling rate 2g of the symmetric cavities compared with the T- and Al1-center optical linewidths across the relevant regimes. $2g/2\pi$ and $2g/\kappa$ are taken from the main text (Fig. 3b); a ratio $2g/\gamma > 1$ places the emitter, together with $2g/\kappa > 1$, in the strong-coupling regime. The homogeneous linewidth of the T center is calculated at 3 to 4 K from the $TX_0$ to $TX_1$ thermal-activation model of Bergeron et al.; for the Al1 center, the value reported at 4.2 K is an upper bound over this range.

| Linewidth regime | Optical linewidth $\gamma/2\pi$ | $2g/\gamma$ | Coupling regime | Reference |
|---|---|---|---|---|
| **T center ($2g/2\pi$ = 675 MHz; $2g/\kappa$ = 6.0)** | | | | |
| Isolated emitter (transform limit) | 0.17 MHz | ~3970 | cavity-limited | [6] |
| Homogeneous, 3 to 4 K | 0.039 to 0.18 GHz | 3.7 to 17 | cavity-limited | [6] |
| Charge-stabilized, sub-2 K | 0.11 to 0.13 GHz | 5.2 to 6.1 | cavity-limited | [14,15] |
| Spectral-diffusion broadened | 1.1 to 4.6 GHz | 0.15 to 0.61 | emitter-limited | [13,16] |
| **Al1 center ($2g/2\pi$ = 1240 MHz; $2g/\kappa$ = 20)** | | | | |
| Isolated emitter (transform limit) | 1.2 MHz | ~1030 | cavity-limited | [9] |
| Homogeneous, 3 to 4 K | ≤ 47 MHz | ≥ 26.4 | cavity-limited | [9] |
| Spectral-diffusion broadened | 10.85 GHz | ~0.11 | emitter-limited | [9] |

Two conclusions follow. First, at the radiative limit both cavities are firmly cavity-limited, with $2g/\gamma$ of order $10^3$, so it is the cavity design, and not the coherent coupling rate, that sets the approach to strong coupling. Second, the two centers differ at realistic operating conditions. The Al1 center remains cavity-limited even at its measured 4.2 K homogeneous linewidth (47 MHz, $2g/\gamma \approx 26$), reflecting a homogeneous linewidth several times narrower than that of the T center near 4 K. The T center is instead more sensitive to temperature: $TX_0$–$TX_1$ phonon mixing broadens its homogeneous linewidth from about 40 MHz at 3 K to about 180 MHz at 4 K ($2g/\gamma \approx 17$ to 3.7), so the T center remains cavity-limited at 3 to 4 K, with a margin that shrinks rapidly as the temperature rises, provided that spectral diffusion is also suppressed.

Under the optical linewidths typically measured for single emitters, both centers are emitter-limited ($2g/\gamma^* \approx 0.11$ to 0.61), owing to inhomogeneous spectral diffusion of order 1 to 10 GHz. This broadening originates in the charge and strain environment, namely implantation-induced lattice damage for the Al1 center and charge-noise-induced spectral wandering for the T center, rather than in the cavity, and it is being actively reduced. Charge stabilization has narrowed the T-center optical linewidth to about 0.11 to 0.13 GHz [14,15], using a heralded resonance-check scheme and unheralded above-band charge reset in a p-i-n waveguide, with further narrowing demonstrated by above-band optical control[17] and by surface passivation[18], while the initial Al1 report identifies improved implantation and annealing as the route to better spectral stability [9]. At the demonstrated narrowed T-center linewidth, $2g/\gamma$ rises above unity ($2g/\gamma \approx 5.2$ to 6.1). The emitter-limited regime is therefore a material-quality and charge-environment limitation that is being closed experimentally, and not a limitation of the present cavity designs.

Table S7 extends this comparison to the four cavities and to the total loss $\kappa + \gamma^*$. The asymmetric Al1 cavity, the only asymmetric design with $2g/\kappa > 1$ (1.8), remains above unity in the total-loss convention for an isolated emitter ($2g/(\kappa + \gamma^*) \approx 1.8$) and at its homogeneous linewidth (≥ 1.65). Under spectral-diffusion broadening, however, $2g/(\kappa + \gamma^*)$ falls below unity for all four cavities (0.10 to 0.56), which then operate in the weak-coupling (Purcell) regime.

**Table S7.** Coupling-to-loss ratios of the four cavities under the three conventions. $\kappa/2\pi$ is the loaded cavity linewidth ($\kappa = \omega/Q$) and $2g/2\pi$ the coherent coupling rate (Equation (S12)), both simulated; they are given in the header row of each cavity together with $2g/\kappa$, which does not depend on the emitter linewidth. $\gamma^*/2\pi$ is the full optical linewidth of the emitter in each regime, with the values and references of Table S6. $2g/\kappa$ compares the coherent coupling with the cavity loss alone (main-text convention, Fig. 3b), $2g/\gamma^*$ with the emitter linewidth alone (Table S6), and $2g/(\kappa + \gamma^*)$ with the total loss, the most conservative of the three; a ratio above unity indicates that coherent exchange exceeds the corresponding loss.

| Linewidth regime | Optical linewidth $\gamma^*/2\pi$ | $2g/\gamma^*$ | $2g/(\kappa + \gamma^*)$ |
|---|---|---|---|
| **T symmetric (κ/2π = 112 MHz; 2g/2π = 675 MHz; 2g/κ = 6.0)** | | | |
| Isolated emitter (transform limit) | 0.17 MHz | ~3970 | ~6.0 |
| Homogeneous, 3 to 4 K | 0.039 to 0.18 GHz | 3.7 to 17 | 2.3 to 4.5 |
| Charge-stabilized, sub-2 K | 0.11 to 0.13 GHz | 5.2 to 6.1 | 2.8 to 3.0 |
| Spectral-diffusion broadened | 1.1 to 4.6 GHz | 0.15 to 0.61 | 0.14 to 0.56 |
| **T asymmetric (κ/2π = 1947 MHz; 2g/2π = 671 MHz; 2g/κ = 0.34)** | | | |
| Isolated emitter (transform limit) | 0.17 MHz | ~3950 | ~0.34 |
| Homogeneous, 3 to 4 K | 0.039 to 0.18 GHz | 3.7 to 17 | 0.31 to 0.34 |
| Charge-stabilized, sub-2 K | 0.11 to 0.13 GHz | 5.2 to 6.1 | 0.32 to 0.33 |
| Spectral-diffusion broadened | 1.1 to 4.6 GHz | 0.15 to 0.61 | 0.10 to 0.22 |
| **Al1 symmetric (κ/2π = 62 MHz; 2g/2π = 1240 MHz; 2g/κ = 20)** | | | |
| Isolated emitter (transform limit) | 1.2 MHz | ~1030 | ~20 |
| Homogeneous, 3 to 4 K | ≤ 47 MHz | ≥ 26.4 | ≥ 11.4 |
| Spectral-diffusion broadened | 10.85 GHz | ~0.11 | ~0.11 |
| **Al1 asymmetric (κ/2π = 704 MHz; 2g/2π = 1236 MHz; 2g/κ = 1.8)** | | | |
| Isolated emitter (transform limit) | 1.2 MHz | ~1030 | ~1.8 |
| Homogeneous, 3 to 4 K | ≤ 47 MHz | ≥ 26.3 | ≥ 1.65 |
| Spectral-diffusion broadened | 10.85 GHz | ~0.11 | ~0.11 |